\documentclass[a4paper,11pt]{article}
\pdfoutput=1 

\usepackage{jinstpub} 
\usepackage{graphicx}
\usepackage{subfig}
\usepackage{lineno}
\usepackage{multirow}

\title{The ENUBET positron tagger prototype: construction and testbeam performance}

\newcommand{\itns}        [3]  {{\emph{IEEE Trans. Nucl. Sci.}}~{\bf #1} (#2) #3}

\author[a,b]{F.~Acerbi,}
\author[e]{M.~Bonesini,}
\author[f]{F.~Bramati,}
\author[e,f]{A.~Branca,}
\author[e,f]{C.~Brizzolari,}
\author[c]{G.~Brunetti,}
\author[e,p]{S.~Capelli,}
\author[d,g]{S.~Carturan,}
\author[h]{M.G.~Catanesi,}
\author[i]{S.~Cecchini,}
\author[i]{F.~Cindolo,}
\author[c,d]{G.~Collazuol,}
\author[c]{E.~Conti,}
\author[c]{F.~Dal Corso,}
\author[c,d]{C.~Delogu,}
\author[j,k]{G.~De~Rosa,}
\author[e,f]{A.~Falcone,}
\author[a]{A.~Gola,}
\author[l]{C.~Jollet,}
\author[m]{B.~Kli\u{c}ek,}
\author[n,u,v]{Y.~Kudenko,}
\author[c,d]{M.~Laveder,}
\author[c,d]{A.~Longhin,}
\author[o]{L.~Ludovici,}
\author[e,p]{E.~Lutsenko,}
\author[h,q]{L.~Magaletti,}
\author[i]{G.~Mandrioli,}
\author[i]{A.~Margotti,}
\author[e,p]{V.~Mascagna,}
\author[i]{N.~Mauri,}
\author[e,f]{L.~Meazza,}
\author[l]{A.~Meregaglia,}
\author[c]{M.~Mezzetto,}
\author[t]{A.~Paoloni,}
\author[c,d,r]{M.~Pari,}
\author[e,f,r]{E.~Parozzi,}
\author[i,s]{L.~Pasqualini,}
\author[a]{G.~Paternoster,}
\author[i]{L.~Patrizii,}
\author[i]{M.~Pozzato,}
\author[e,p]{M.~Prest,}
\author[c,d]{F.~Pupilli,}
\author[h]{E.~Radicioni,}
\author[j,k]{C.~Riccio,}
\author[j,k]{A.C.~Ruggeri,}
\author[c,d]{C.~Scian,}
\author[i]{G.~Sirri,}
\author[m]{M.~Stip\u{c}evi\'{c},}
\author[i]{M.~Tenti,}
\author[e,f]{F.~Terranova,}
\author[e,f,1]{M.~Torti\note{Corresponding author.},}
\author[e]{E.~Vallazza,}
\author[t]{L.~Votano}

\affiliation[a]{Fondazione Bruno Kessler (FBK),  Via Sommarive 18 - 38123 Povo (TN), IT}
\affiliation[b]{INFN-TIFPA, Università di Trento, Via Sommarive 14 - 38123 Povo (TN), IT}
\affiliation[c]{INFN Sezione di Padova, via Marzolo 8 - 35131 Padova, IT}
\affiliation[d]{Università di Padova, via Marzolo 8 - 35131 Padova, IT}
\affiliation[e]{INFN Sezione di Milano-Bicocca, Piazza della Scienza 3 - 20133 Milano, IT}
\affiliation[f]{Università di Milano-Bicocca, Piazza della Scienza 3 - 20133 Milano, IT}
\affiliation[g]{INFN, Laboratori Nazionali di Legnaro, Viale dell'Università 2 - 35020 Legnaro (PD), IT}
\affiliation[h]{INFN Sezione di Bari, Via Giovanni Amendola 173 - 70126 Bari, IT}
\affiliation[i]{INFN Sezione di Bologna, viale Berti-Pichat 6/2 - 40127 Bologna, IT}
\affiliation[j]{INFN, Sezione di Napoli, Strada Comunale Cinthia - 80126 Napoli, IT}
\affiliation[k]{Università ``Federico II'' di Napoli, Corso Umberto I 40 - 80138 Napoli, IT}
\affiliation[l]{CENBG, Universitè de Bordeaux, CNRS/IN2P3, 33175 Gradignan, FR}
\affiliation[m]{Center of Excellence for Advanced Materials and Sensing Devices, Ruđer Bošković
Institute, HR-10000 Zagreb, HR}
\affiliation[n]{
Institute for Nuclear Research of the Russian Academy of Sciences, 117312
Moscow, RU}
\affiliation[o]{INFN Sezione di Roma 1, Piazzale A. Moro 2, 00185 Rome, IT}
\affiliation[p]{Università degli Studi dell'Insubria, Via Valleggio 11 - 22100 Como, IT}
\affiliation[q]{Università degli Studi di Bari, Via Giovanni Amendola 173 - 70126 Bari, IT}
\affiliation[r]{CERN, Esplanade des particules - 1211 Genève 23, CH}
\affiliation[s]{Università degli Studi di Bologna, viale Berti-Pichat 6/2 - 40127 Bologna, IT}
\affiliation[t]{INFN, Laboratori Nazionali di Frascati, via Fermi 40 - 00044 Frascati (Rome), Italy}
\affiliation[u]{National Research Nuclear University “MEPhI”, 115409 Moscow, Russia}
\affiliation[v]{Moscow Institute of Physics and Technology, 141701 Moscow region, Russia}

\emailAdd{marta.torti@mib.infn.it}

\abstract{A prototype for the instrumented decay tunnel of ENUBET  was tested in 2018 at the CERN East Area facility with charged particles up to 5~GeV. This detector is a longitudinal sampling calorimeter with lateral scintillation light readout. The calorimeter was equipped by an additional ``$t_0$-layer'' for timing and photon discrimination. 
The performance of this detector in terms of electron energy resolution, linearity, response to muons and hadron showers are presented in this paper and compared with simulation. The $t_0$-layer was studied both in standalone mode using pion charge exchange and in combined mode with the calorimeter to assess the light yield and the 1~mip/2~mip separation capability. 
We demonstrate that this system fulfills the requirements for neutrino physics applications and discuss performance and additional improvements.}


\begin{document}
\maketitle
\flushbottom

\section{Introduction}
\label{sec:intro}

Monitored neutrino beams \cite{Longhin:2014yta} are highly controlled sources of neutrinos at the GeV scale. They represent the ideal facilities for a new generation of experiments to measure neutrino cross sections of relevance for oscillation studies at per cent level. In particular, the ERC ENUBET project \cite{enubet_eoi,enubet_erc,enubet_proposal} is aimed at designing the first monitored beam tagging large angle positrons from the three body decay of charged kaons ($K_{e3}: \ K^+ \rightarrow e^+ \pi^0 \nu_e $) and thus providing a pure source of $\nu_e$ where the flux is measured with a precision of $<1$\%.    
One of the key challenges of ENUBET is to devise a compact, radiation-hard, efficient and cost-effective instrumentation for the decay tunnel, whose requirements are detailed in Sec.~\ref{sec:requirements}. The ENUBET ``positron tagger'' must be  capable to identify electrons and muons in the  neutrino decay tunnel located after a narrow band secondary transfer line.  
In 2017-2018 we demonstrated that the ENUBET requirements can be achieved using an iron-scintillator calorimeter whose basic unit is an Ultra Compact Module (UCM) sampling e.m. and hadronic showers every 4.3 radiation lengths ($X_0$) or, equivalently, 0.45 interaction lengths ($\lambda_0$) \cite{Ballerini:2018hus}. Scintillation light produced by five 0.5~cm thick tiles is transported by wavelength-shifter (WLS) fibers crossing the 1.5~cm thick iron tiles (''shashlik'' light readout \cite{Berra:2016thx,test_12module,Acerbi:2020itd}) toward 1~mm$^2$ Silicon Photomultipliers (SiPMs) located in the back of the module and, hence, embedded in the bulk of the calorimeter.
Irradiation tests \cite{sipmirr} performed in 2018 demonstrate that the most critical component of the UCM are the SiPMs, which are exposed to fast neutrons produced by hadronic showers. The ENUBET UCMs are able to stand up to $\mathcal{O}(10^{11})$~n/cm$^2$, which is sufficient for $\nu_e$ cross section measurements with a statistical uncertainty of 1\% employing  neutrino detectors of the same size of ICARUS at Fermilab \cite{Antonello:2015lea} or the ProtoDUNEs at CERN \cite{Abi:2017aow,Agostino:2014qoa}. The UCM, however, has two drawbacks: SiPMs are inaccessible for maintenance during data taking and fluxes exceeding $\mathcal{O}(10^{12})$~n/cm$^2$ may compromise the sensitivity of the UCM to muons if data taking is significantly extended or the average beam power is increased well above the ENUBET baseline design.

\begin{figure} [!htb]
\centering
  \includegraphics[width=0.6\linewidth]{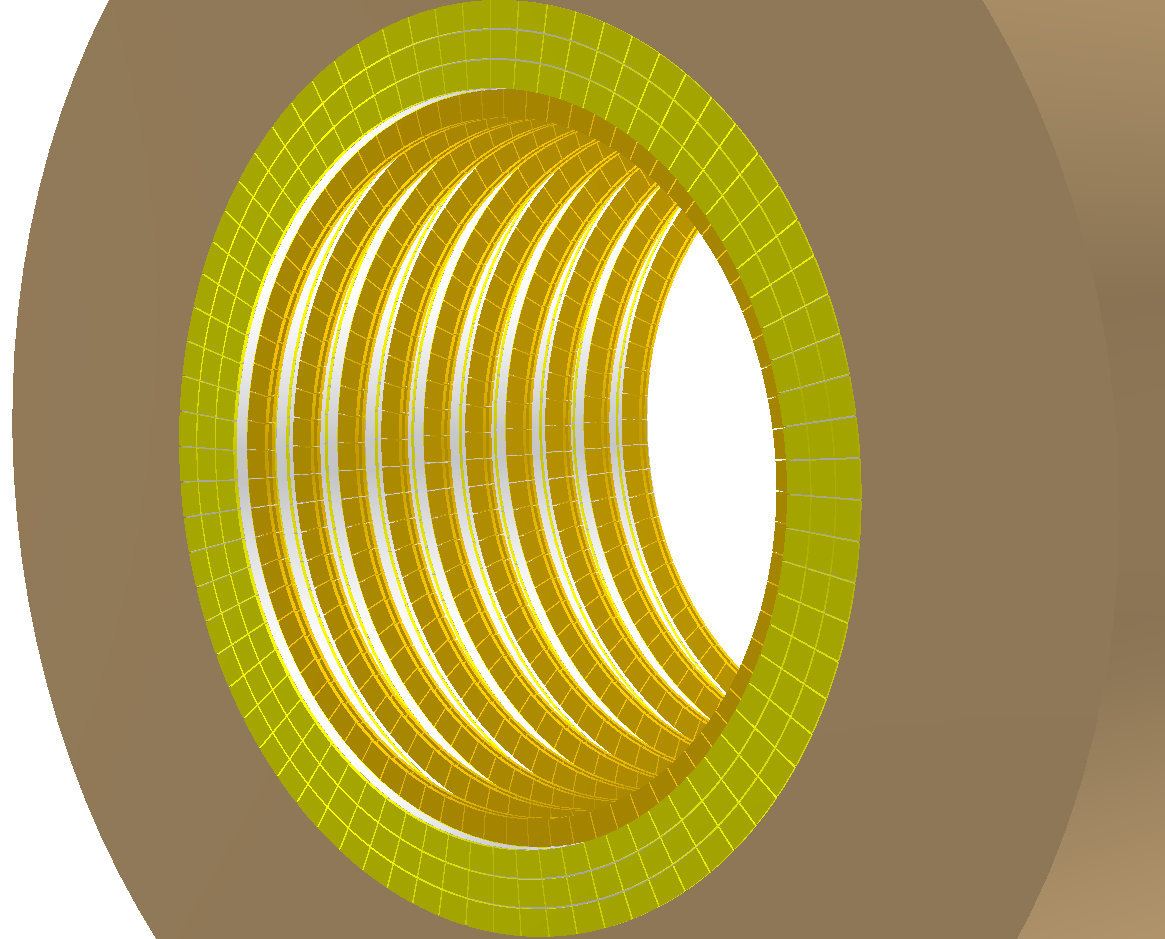}
  \caption[]{Schematics of the ENUBET instrumented decay tunnel. The three layers of modules of the calorimeter (light green) constitute the inner wall of the tunnel.  The rings of the scintillator tiles of the photon veto (yellow) are located just below the modules. The length of each module is 10~cm and the tile doublets of the photon veto are installed every 7~cm. In the lateral readout option, the optical fibers (not shown) bring the light in the radial direction toward the outer part of the tunnel (light brown) where the SiPMs (not shown) are positioned.  }
  \label{fig:tagger}
\end{figure}

The tolerable beam power can be increased by a factor of about 18 (see Sec.~\ref{sec:neutron_red}) positioning the SiPMs above the calorimeter 
and on top of a 30~cm Borated polyehtylene shield and transporting the light from the module to the top of the detector by WLS fibers running along one of the lateral side of the tiles.   
This setup replaces the shashlik-based UCM with a lateral readout compact module (LCM) where the light of all fibers belonging to a module is recorded by a single $4\times 4$~mm$^2$ SiPM. The number of SiPMs is thus equal to the number of modules and they can be accessed during data taking from the outer part of the decay tunnel. The drawback of this setup is an increased complexity in mechanical installation and a slight reduction of the light yield due to the longer fiber length. 

These challenges were addressed by ENUBET in 2018-2019 by constructing a prototype calorimeter whose size is similar to the UCM-based detector \cite{Ballerini:2018hus} but it is assembled from LCMs. In addition, this prototype was equipped with a $t_0$-layer built with a technology very similar to the LCM: tile doublets of plastic scintillator laterally readout by WLS fibers that are connected to the SiPMs. In this paper, we describe the design and construction of the lateral readout calorimeter  (Sec.~\ref{subsec:des_calo}) and the $t_0$-layer (Sec.~\ref{subsec:des_t0layer}). The positron tagger, i.e. the calorimeter equipped with the $t_0$-layer, was tested at the CERN East Experimental Area in fall 2018 (see Sec.~\ref{sec:experiment}): we show the performance of the detector in a mixed beam of electrons, muons and hadrons  in the energy range of interest for monitored neutrino beams in Sec.~\ref{sec:mip}, \ref{sec:electrons} and \ref{sec:hadrons}. The performances of the $t_0$-layer are detailed in Sec.~\ref{sec:t0layer}. 

\section{Requirements}
\label{sec:requirements}

ENUBET will provide the most sophisticated diagnostics ever conceived for neutrino beams to control the $\nu_e$ and $\nu_\mu$ neutrino flux at source. This is motivated by the uncertainties on the neutrino cross sections at the GeV scale that limit the physics reach of future neutrino oscillation experiments (in particular, DUNE and Hyper-Kamiokande). In previous neutrino cross-section experiments, the measurement systematics are completely dominated by the uncertainties on the flux, which will be overcome by ENUBET. 

For a transfer line selecting kaons at 8.5~GeV/c, like the one envisaged for ENUBET, the positrons from $K_{e3}$ reaching the instrumented walls of the tunnel at 1~m from the beam axis span an energy range between 1 and 3 GeV. The mean energy of the positron is $\sim$1.6~GeV and the mean angle is $\sim$125~mrad. Since the positrons produced per spill exceed $10^{7}$, statistical error is always negligible and we aim at recording a (prescaled) subsample of minimum bias events for monitoring purposes.  The main background consists of charged pions from the other decay modes of the kaons and from the off-momentum beam halo transported at the entrance of the tunnel. In addition, the instrumentation must be able to suppress muons from decays along the beamline (halo muons) and photons from tertiary e.m. showers and $\pi^0$.
An overall positron efficiency of 20\% or more with a signal-to-noise ratio $>1$ is sufficient to predict the $\nu_e$ flux at per-cent level.
The instrumentation must be cost-effective and reliable and should be placed around the wall of the decay tunnel covering a significant fraction of this 40~m long tunnel. Sampling calorimeter with longitudinal segmentation read out by WLS fibers and compact solid-state photosensors fulfill these requirements and represent the technology of choice for ENUBET.
A full simulation of the ENUBET beamline performed in 2016-2020 \cite{pupilli, longhin_SPSC} indicates that an appropriate $e^+/\pi^+$ separation can be achieved by longitudinally segmented sampling calorimeters with an e.m. energy resolution $<25\%/\sqrt{E(\mathrm{GeV})}$ in the  range of interest for ENUBET (1-3~GeV). Charged pions are separated by positrons (or background electrons) employing the energy deposition pattern in the longitudinal modules of the calorimeter. Positron identification has been simulated starting from particles transported by the ENUBET beamline at the entrance of the decay tunnel. The ENUBET GEANT4 simulation includes particle tracking and detector response and the full particle identification (PID) chain from the event builder to the positron identification. The PID is based on a Multivariate Data Analysis (TMVA)~\cite{Meregaglia:2016vxf,Hocker:2007ht} that employs 13 variables constructed from the energy deposit in each module. The latest results provide a signal-to-noise ratio of $\sim$1.6 for a positron efficiency\footnote{including the geometrical acceptance that amounts to $\sim$53\%} of $\sim$24\% using the calorimeter described in this paper.

The particles impinge in the calorimeter from the innermost part of the tunnel in an angular range between 10 and 200~mrad and the simulation accounts for energy losses due to albedo, i.e. parts of the shower that bounce back into the tunnel. The average particle rate in the calorimeter
along the tunnel length is $\sim$600~kHz and is dominated by kaon decay products and muons from the beam halo. Pile-up suppression requires a recovery time of about 20~ns which is achieved both by fast recovery devices and by the analysis of the SiPM waveform recorded during the spill. 
The ENUBET analysis includes pile-up effects simulated through GEANT4~\cite{Agostinelli:2002hh,Allison:2006ve,Allison:2016lfl} by the time distribution of the particles entering the decay tunnel.
The response of the SiPM is simulated at hardware level using GosSiP~\cite{Eckert:2012yr} and dedicated algorithms to further disentangling pile-up are being developed. 

Longitudinal segmentation does not provide separation between positrons and photons. Photons originate from showers produced in the transfer line and from $K^+$ decays that produce neutral pions (in particular $K^+ \rightarrow \pi^0 \pi^+$ decays). The positron tagger (see Fig.~\ref{fig:tagger}) must hence include a photon veto, which also provides the timing of the event (``$t_0$-layer'').
The $t_0$-layer must provide absolute timing of the events with a precision $<2$~ns, an efficiency for a single minimum ionizing particle (mip) $>90$\%
and 1 mip/2 mip discrimination capability to reject the small fraction of photons that converts inside a tile of the photon veto.

Finally, the ENUBET instrumentation must be radiation tolerant both for ionizing and non ionizing doses. Doses depends on the quality of the beam and the duration of the run. A full dose assessment was performed using the FLUKA 2011 code~\cite{fluka1,fluka2}
and demonstrated that the only critical component are the SiPMs if they are embedded in the core of the calorimeters. 
These studies (see Sec.~\ref{sec:intro} and \ref{sec:neutron_red}) motivated the design of the lateral readout scheme. In particular, the neutron fluence on the inner surface of the calorimeter for a run that collects $10^4$ $\nu_e$ CC events at the neutrino detector is $\sim 2 \times 10^{11}$ n/cm$^2$ (1~MeV equivalent) but lower irradiation levels are highly advisable especially to prevent the dark count rate to disrupt the identification of minimum ionizing particles~\cite{sipmirr}. Mip identification is useful for self-monitoring possible drifts of the LCM response in the course of the run. Muon identification allows to exploit additional $K$ decay channels ($K^+ \rightarrow \mu^+ \nu_\mu$) and constrain the distribution of halo muons from the transfer line.  

The results presented in this paper validate a significant fraction of this simulation for the LCM-based instrumentation: the energy response of the electron at various angles, the data/MC agreement for the response to the mip, the longitudinal profile of charged pions and, finally, the performance of the $t_0$-layer. 
Further details and opportunities offered by NP06/ENUBET
(muon monitoring after the hadron dump for the $\nu_\mu$, a priori measurement of the $\nu_\mu$ at source, tagged neutrino beams etc.) are summarized in Ref.~\cite{longhin_SPSC}.

\section{Layout and construction of the positron tagger}
\label{sec:detector}

The prototype of the ENUBET positron tagger is composed by the longitudinal segmented calorimeter based on the LCM and the $t_0$-layer. 

\subsection{Description of the calorimeter } 
 \label{subsec:des_calo}
The construction of the prototype calorimeter was performed in two steps. In April 2018, we built a 
setup (``Module 1'') composed of 18~LCMs to validate the production and mounting procedure and estimate the light yield. In summer 2018, we added two additional modules: ``Module 2'', identical to Module 1 i.e. composed by 18~LCMs, and ``Module 3'' made of 48~LCMs. 

\begin{figure} [ht]
\centering
  \includegraphics[width=0.7\linewidth]{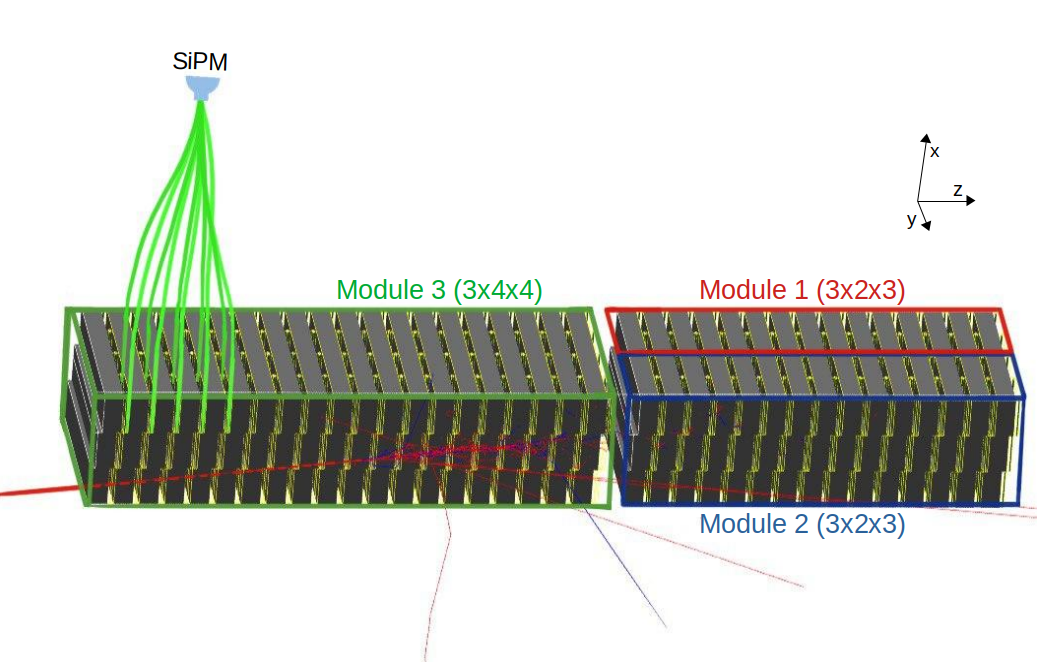}
  \caption[]{Layout of the three Modules of the calorimeter. The GEANT4 simulation shows a 3 GeV electron impinging at 200 mrad the side under the calorimeter. The numbers in parenthesis correspond to the LCMs of the Modules. The $z$ direction corresponds to the axis of the tunnel. The $x$ axis correspond to the radial direction of the tunnel while $y$ cover a fraction of the azimuthal direction of the tunnel.}
  \label{fig:scheme_modules}
\end{figure}

\begin{figure} [!htb]
\centering
  \includegraphics[width=0.5\linewidth]{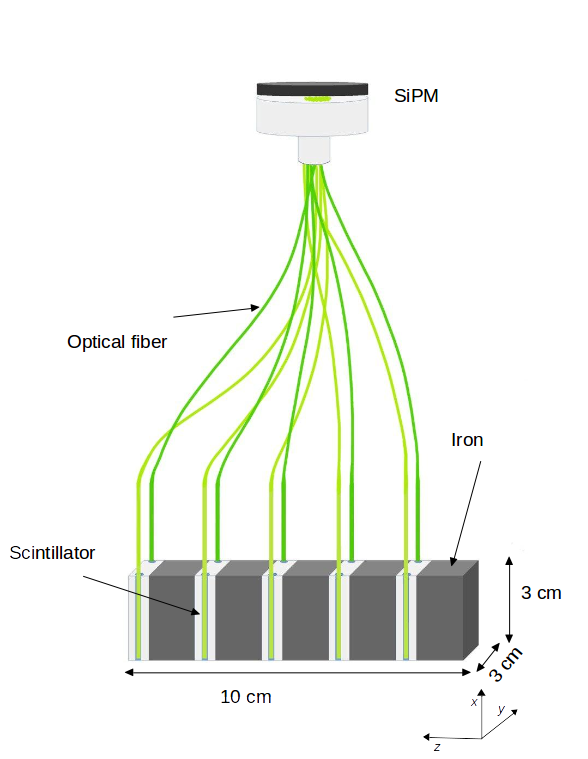}
  \caption[]{Layout of one LCM. The axis definition is the same as for Fig.~\ref{fig:scheme_modules}}
  \label{fig:lcm_schema}
\end{figure}

The prototype tested in fall 2018  is thus composed of three parts (see Fig.~\ref{fig:scheme_modules} and Tab.~\ref{tab:components}). Module 1 and 2 have 
2 columns of LCMs in the axis that corresponds to the azimuthal direction in Fig.~\ref{fig:tagger} ($y$-axis in Tab.~\ref{tab:components}). They have 3 layers in the transverse beam plane, i.e. in the radial direction of Fig.~\ref{fig:tagger} ($x$-axis in Tab.~\ref{tab:components}) and 3 planes in the longitudinal direction along the beam axis ($z$-axis in Tab.~\ref{tab:components}). Each LCM is composed by  1.5~cm thick iron slabs interleaved with 0.5~cm plastic scintillators (Eljen EJ-204 \cite{eljen}), while each scintillator tile and iron slab has a cross section of 3×3~cm$^2$, as sketched in Fig.\ref{fig:lcm_schema}. The scintillator tiles are painted with a diffusive TiO$_2$-based coating and read by two WLS Saint Gobain BCF92 fibers~\cite{saintgobain} with a diameter of 1~mm each glued to the sides. In addition, a Mylar\textsuperscript \textregistered  \ (BoPET) foil is placed between the two columns. 

\begin{figure} [!htb]
\centering
  \includegraphics[width=0.5\linewidth]{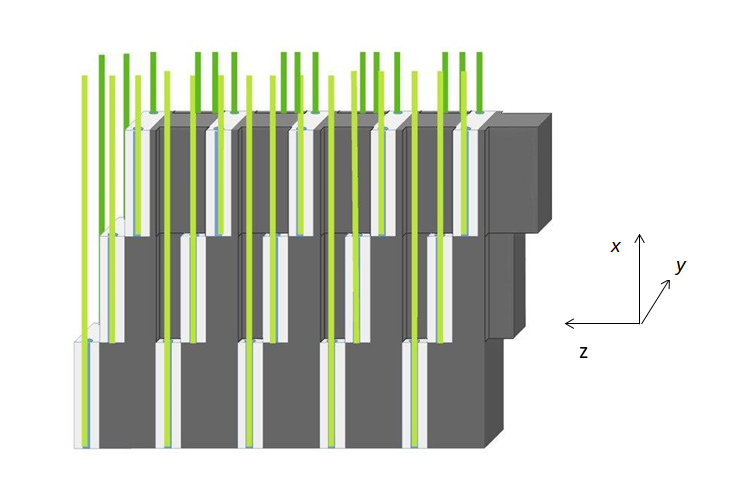}
  \caption[]{Layout of the fibers running toward the SiPMs through several LCMs by the iron grooves. The axis definition is the same as for Fig.~\ref{fig:scheme_modules}}
  \label{fig:fibers_proto}
\end{figure}

\begin{figure} [!htb]
\centering
  \includegraphics[width=0.54\linewidth]{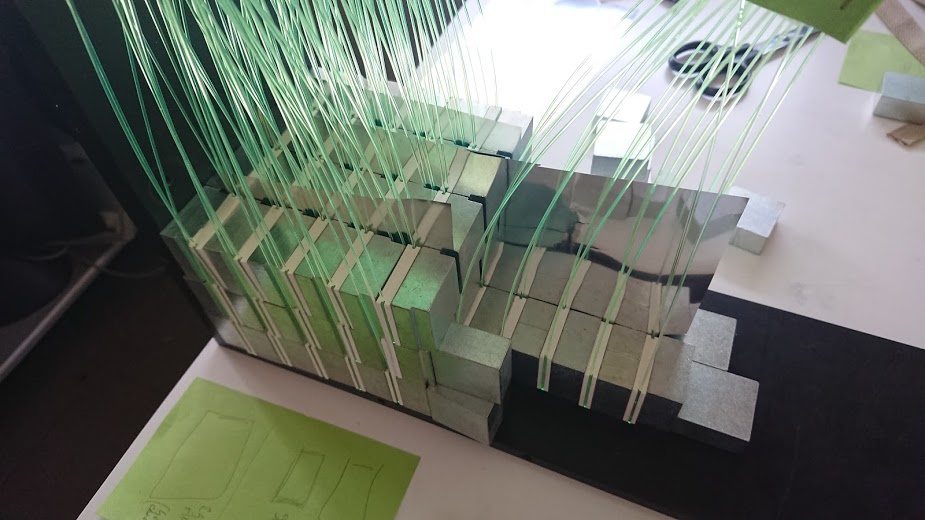}
    \includegraphics[width=0.41\linewidth]{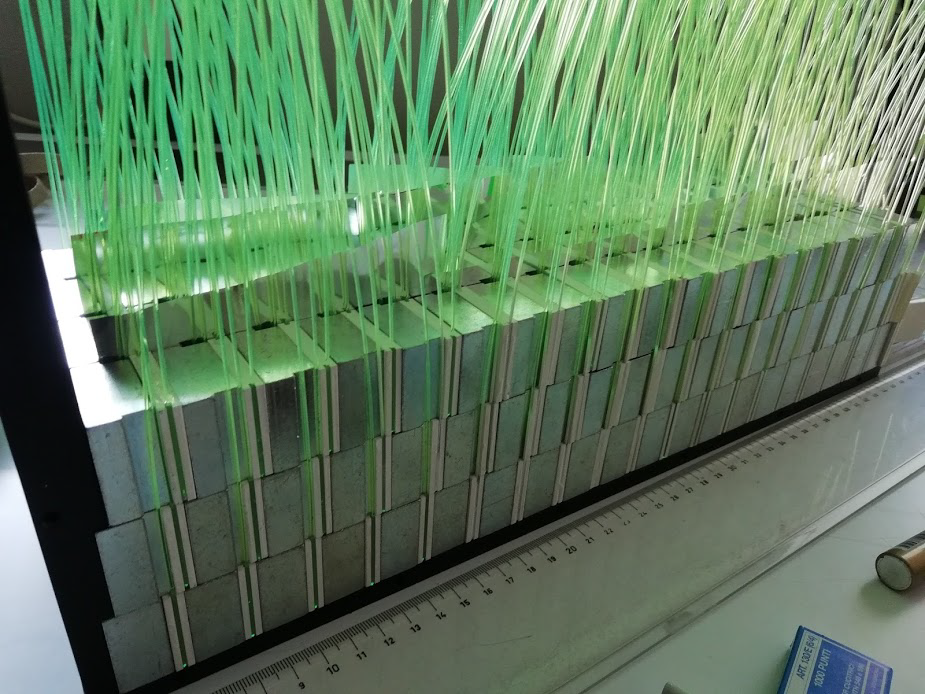}
  \caption[]{Mounting of one of the Modules before the bundling of the fibers and the installation of the SiPMs.}
  \label{fig:mounting}
\end{figure}

The longitudinal planes are shifted by 3.5~mm with respect to each other, to allow for the extraction of the fibers from the bulk of the calorimeter. 
This arrangement (see Fig.~\ref{fig:fibers_proto}), in which each fiber is coupled only to one scintillator tile and does not collect light from the above planes, ensures optical insulation among LCMs. 
For the same reason, iron tiles are grooved on both sides to make room for the passage of the fibers (see Fig.~\ref{fig:mounting}). The grooves are 3~mm and 7~mm wide and 1.5~mm deep. The only difference between the two small Modules is the dimension of the scintillator: the tiles employed in one of them are 0.2~mm thicker than the other ones since they were procured by different Eljen production batches. Hence the LCM length is 10.1~cm instead of 10.0~cm.

\begin{figure} [ht]
\centering
\includegraphics[width=\linewidth]{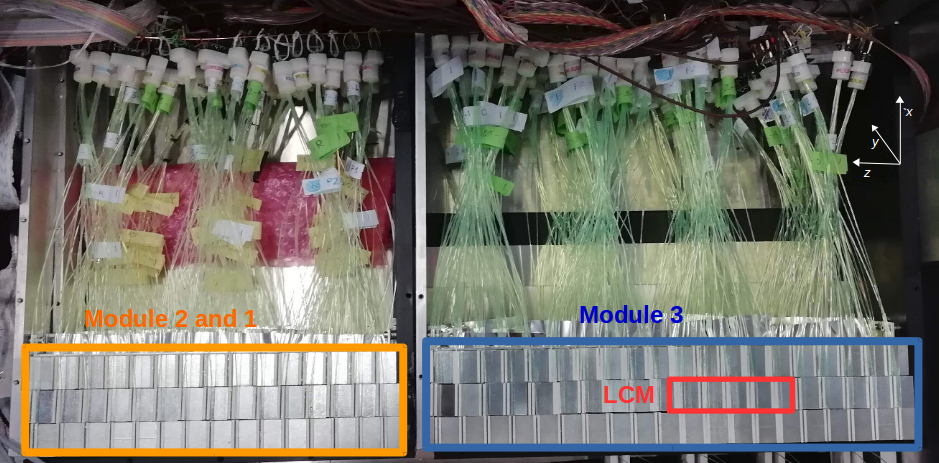}
  \caption[]{The full calorimeter: Module 3 (blue box) on the right, Module 2 (orange box) on the left, Module 1 behind Module 2 (not visible). The red box points out one LCM in the prototype.  
  During the beam-test, the charged particle beam was impinging from the right and Module 1+2 were used for shower containment.}
  \label{fig:prototype}
\end{figure}

Module 3 is made of 3$\times$4$\times$4~modules for a total of 48 LCMs: 3 radial layers, 4 horizontal columns and 4 longitudinal (i.e. along the direction of the beam axis) planes. The first three planes in the longitudinal direction are read by Kuraray Y11 fibers \cite{kuraray}, while the fourth is equipped with Saint Gobain BCF92 fibers. The LCMs are housed in custom PVC/Aluminium boxes and  Mylar foils are positioned among the four columns.
The overall calorimeter -- see Fig.~\ref{fig:prototype} -- thus consists of 84 LCMs in a 3$\times$4$\times$7~structure. The length of the WLS fibers is 30~cm, i.e. three times longer that the previous ENUBET shashlik UCM.
The 10 WLS fibers of a single LCM (corresponding to 5 scintillator tiles) were read by 4$\times$4~mm$^2$ SiPMs with 40~$\mu$m cell size produced by Advansid srl~\cite{Advansid}.  Each SiPM has 9340 cells with a fill factor of 60\% and peak PDE at 550~nm. The breakdown voltage is 27~V  
and the bias is the same for all SiPMs and it is distributed by a coaxial cable. During the beam-test all the SiPMs were biased at 32~V and the equalization among LCMs was performed using the response to minimum ionizing particles (see Sec.~\ref{sec:mip}).

\begin{figure} [!htb]
\centering
  \includegraphics[width=0.3\linewidth]{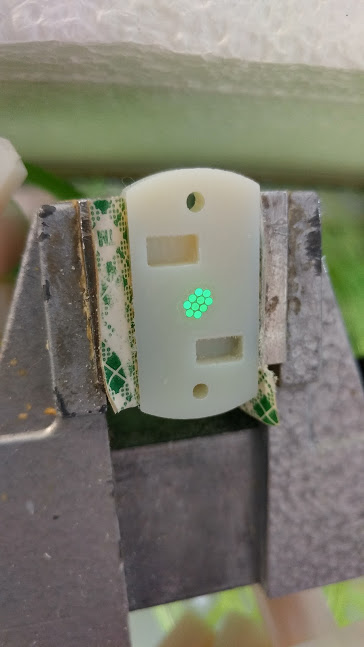}
    \includegraphics[width=0.6\linewidth]{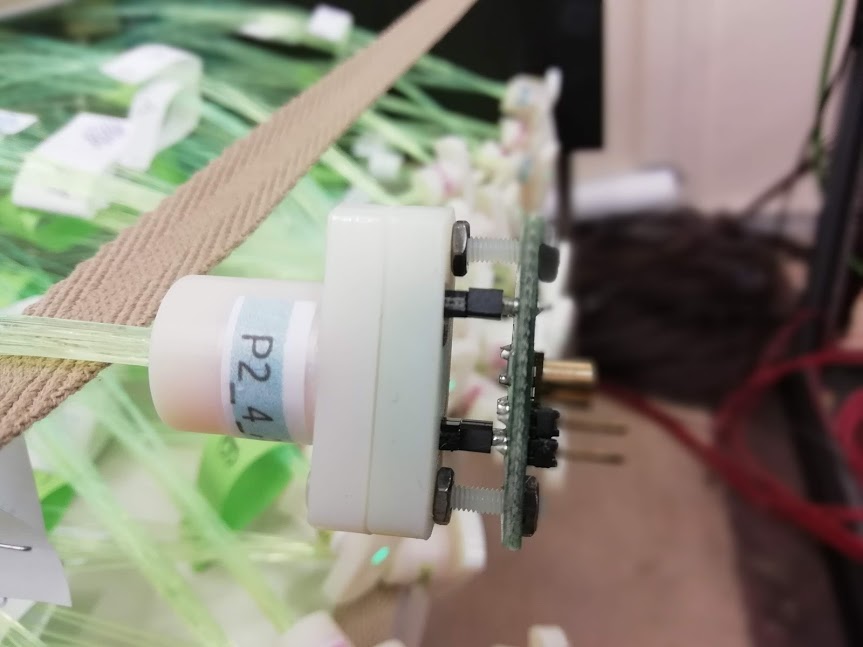} 
  \caption[]{Left: fiber connector close up. Right: detail of the SiPM's connector.}
  \label{fig:connector}
\end{figure}

The signals from the LCMs are recorded by a set of 8 channel v1720 CAEN~\cite{caen} digitizers (12~bit, 250 MS/s).
All waveforms are recorded by the DAQ. A reduced dataset is produced for the analysis employing a peak finding algorithm on the waveform data~\cite{Berra:2016thx}.
The assembly of Module 2 and 3 was optimized taking advantage of the experience gained during the mounting of Module 1. In particular, we developed special plastic connectors to bundle the fibers prior to the final levering and polishing (see Fig.~\ref{fig:connector}, left). The SiPM is hosted in another plastic connector coupled to the bundle with two M3 Teflon screws that hold also the board bringing the bias to the photosensor and leading the anode signal toward the digitizer by a MCX connector (see Fig.~\ref{fig:connector}, right). All plastic connectors were produced by 3D printing at INFN Padova.  

\begin{table}[]
    \centering
    \begin{tabular}{c|c|c|c}
        Unit & Mechanics & Dimension  & Notes   \\ \hline \hline
        LCM & 
        \begin{tabular}{@{}c@{}} 5 iron tiles \\ 5  EJ-204 scint. tiles \\ 10 WLS fibers ($\oslash\  1$~mm) \\ 1 SiPM ($4 \times 4$~mm$^2$) \end{tabular}
         &  
        \begin{tabular}{@{}c@{}}Fe tile: $3 \times 3$ cm$^2$, 1.5~cm thick \\ Scint. tile: $3 \times 3$ cm$^2$, 0.5~cm thick  \\ overall dim.: $3 \times 3 \times 10$~cm$^3$   \end{tabular}
         & 
\begin{tabular}{@{}c@{}}Scint. thick. \\ in M2: 0.52~cm \end{tabular} \\
        \hline
        M1 & LCMs with BCF92 fibers &
   \begin{tabular}{@{}c@{}} 3 LCMs in x, 2 in y, 3 in z, \\ 
   $9 \times 6 \times 30$~cm$^3$ (x,y,z) \\ total: 18 LCMs \end{tabular}
 &  \begin{tabular}{@{}c@{}}fiber polishing \\ not optimized \end{tabular}
    \\ \hline
        M2 & LCMs with BCF92 fibers &
   \begin{tabular}{@{}c@{}} 3 LCMs in x, 2 in y, 3 in z,\\ 
   $9 \times 6 \times 30.3$~cm$^3$ (x,y,z) \\ total: 18 LCMs \end{tabular}
 & \begin{tabular}{@{}c@{}} fiber polishing \\ optimized \\ \end{tabular} \\ \hline
        M3 & LCMs with Y11 fibers &
   \begin{tabular}{@{}c@{}} 3 LCMs in x, 4 in y, 4 in z, \\ 
   $9 \times 12 \times 40$~cm$^3$ (x,y,z) \\ total: 48 LCMs \end{tabular} & see caption (a)  \\
\hline 
All & &
\begin{tabular}{@{}c@{}}  3 LCMs in x, 4 in y, 7 in z, \\ 
$9 \times 12 \times 70$~cm$^3$ (x,y,z) \\ 
total: 84 LCMs \end{tabular} & \\
\hline
$t_0$-layer &
\begin{tabular}{@{}c@{}} pair of EJ-204 scint tiles \\
 BCF92 WLS fibers \\ (2 fibers per tile) \\ SiPMs: SenSL-J 30020 \\ (1 SiPM per tile) \end{tabular}
& \begin{tabular}{@{}c@{}} tile: $3 \times 3 \times 0.5$~cm$^3$ (x,y,z) \\ distance among tiles: 0.5~cm \\ distance among doublets: 7~cm \end{tabular} & 
\begin{tabular}{@{}c@{}} installed below \\ the calo  \end{tabular}
\\ \hline
\end{tabular}
    \caption{Summary of the detector components. The calorimeter is made of 84 LCMs corresponding to 84 channels (1 SiPM per channel) that sample the e.m. and hadronic showers. All SiPMs are located on the top and all fibers are 40~cm long. The $z$ direction corresponds to the axis of the tunnel. The $x$ axis correspond to the radial direction of the tunnel while $y$ cover a fraction (12 cm) of the azimuthal direction of the tunnel. (a) The LCMs of the last ($4^{th}$) plane in $z$ are read by BCF92 fibers. }
    \label{tab:components}
\end{table}

\subsection{The photon veto}
\label{subsec:des_t0layer}
The photon veto ($t_0$-layer) provides both photon identification capabilities and precise timing of the particles in the instrumented decay tunnel. The requirements to achieve the goals of ENUBET are a photon identification efficiency at 99\% and a time resolution of $\sim$1~ns.

\begin{figure}
\centering
  \includegraphics[width=\linewidth]{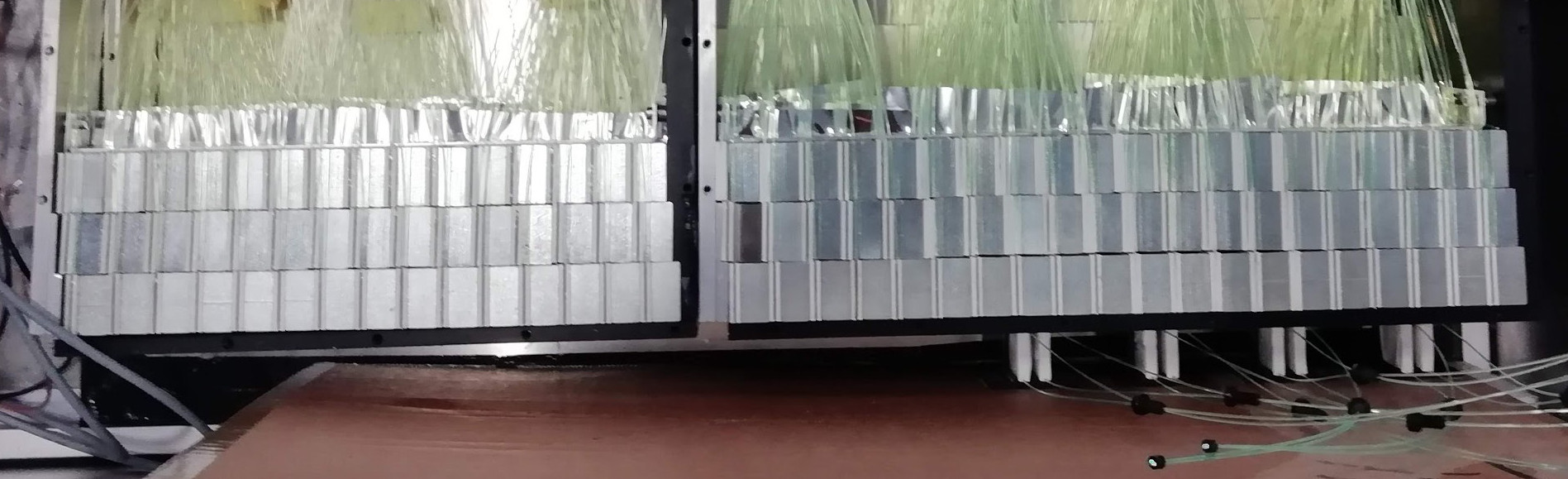}
  \caption[]{Four doublets of the photon veto installed below Module 3 (bottom rightmost part of the figure).}
  \label{fig:veto}
\end{figure}

The $t_0$-layer is composed of doublets of EJ-204 plastic scintillator tiles with a surface of 3$\times$3~cm$^2$ and a thickness of 0.5~cm. The tiles are mounted below the LCMs and positioned every 7~cm (see Fig.~\ref{fig:veto}) so that 
positrons from kaon decays in the ENUBET working condition ($\theta_{e^+} \sim 100$~mrad)
cross five doublets on average. The surface of the tiles is painted with a TiO$_2$ layer. Two 40~cm long BCF92 multi-clad (MC) WLS fibers  are glued to the lateral edges of the tiles with the same optical cement (EJ-500) used for the calorimeter and bundled by a custom connector, which optically couples them to one SenSL-J 30020 SiPM \cite{sensl}. The SenSL SiPMs are equipped with a fast output signal employed for timing applications. The other end of the fibers is covered by a reflective painting. 

Signals from the photon veto are amplified by a custom two stages amplifier ($\sim$130 amplification factor) with a bandwidth up to 500 MHz. Both the anode and the fast output of the SiPMs are acquired by a 10 bit CAEN V1751 digitizer with 2~GS/s resolution through a VME-based DAQ.

\section{Neutron reduction studies}
\label{sec:neutron_red}

We have set up a detailed simulation of a tentative beamline for the ENUBET facility
in the FLUKA framework. The goal is to estimate the ionizing doses and neutron fluences for all the elements of the beamline and, especially, for the decay pipe where the positron instrumentation is located. 

Such layout of the beamline has been defined by a dedicated study performed
with the TRANSPORT~\cite{transport} and G4BeamLine~\cite{g4beamline} codes. The transfer line from the target to the decay tunnel produces an intense and collimated
hadron beam with low levels of stray particles. The optimization performed with G4Beamline minimizes the length of the transfer line to reduce losses from kaon decays occurring before the entrance of the decay tunnel. 

\begin{figure}
\centering
\includegraphics[width=\linewidth]{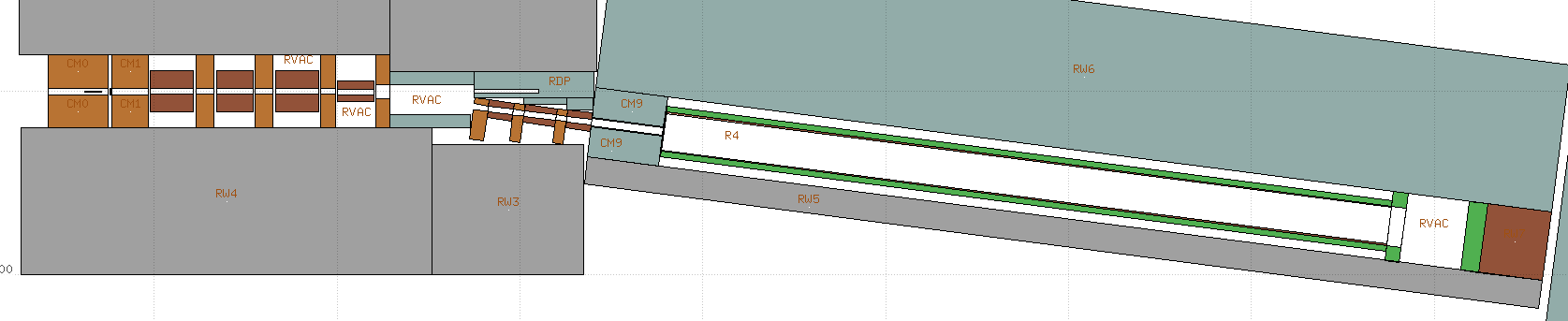}
  \caption[]{Layout of the hadronic beamline modeled with FLUKA. The regions
  in green are those composed of borated polyethylene.}
  \label{fig:FLUKAlayout}
\end{figure}

The resulting system is shown in Fig.~\ref{fig:FLUKAlayout}. In this scheme
the beamline consists of an ``on-axis'' quadrupole triplet followed by a single dipole and an ``off-axis''
quadrupole triplet. The overall bending angle of the resulting
neutrino beam with respect to the proton axis is $\sim$~7.4$^\circ$.
Quadrupoles and dipoles were
dimensioned to achieve a collimated beam of
pions and kaons at an average momentum of 8.5~GeV/c and a momentum
bite of 5-10\% with the shortest possible length to avoid losing too
many neutrinos from early decays of kaons ($\beta\gamma c \tau
\sim$~63~m at 8.5~GeV/c). The optimization of the position and size of the
proton dump is in progress. 

While implementing the beam optics in FLUKA several optimizations were
performed mainly in terms of collimators and shielding. In particular a
Tungsten plug of 4~m in length was added in front of the decay pipe to
protect the calorimeter from background particles thanks to the large
stopping power of this dense material. 

\begin{figure}
  \includegraphics[width=0.435\linewidth]{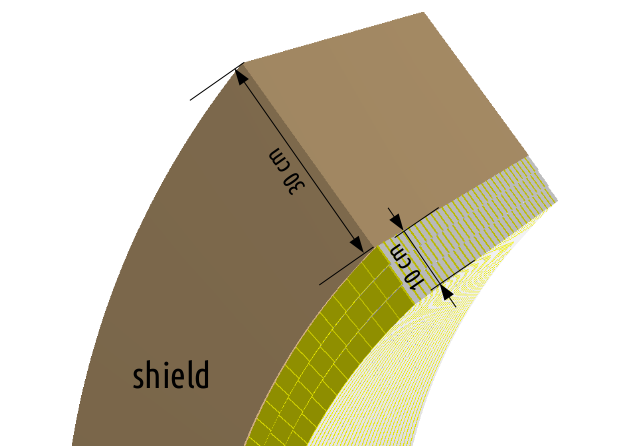}
  \includegraphics[width=0.565\linewidth]{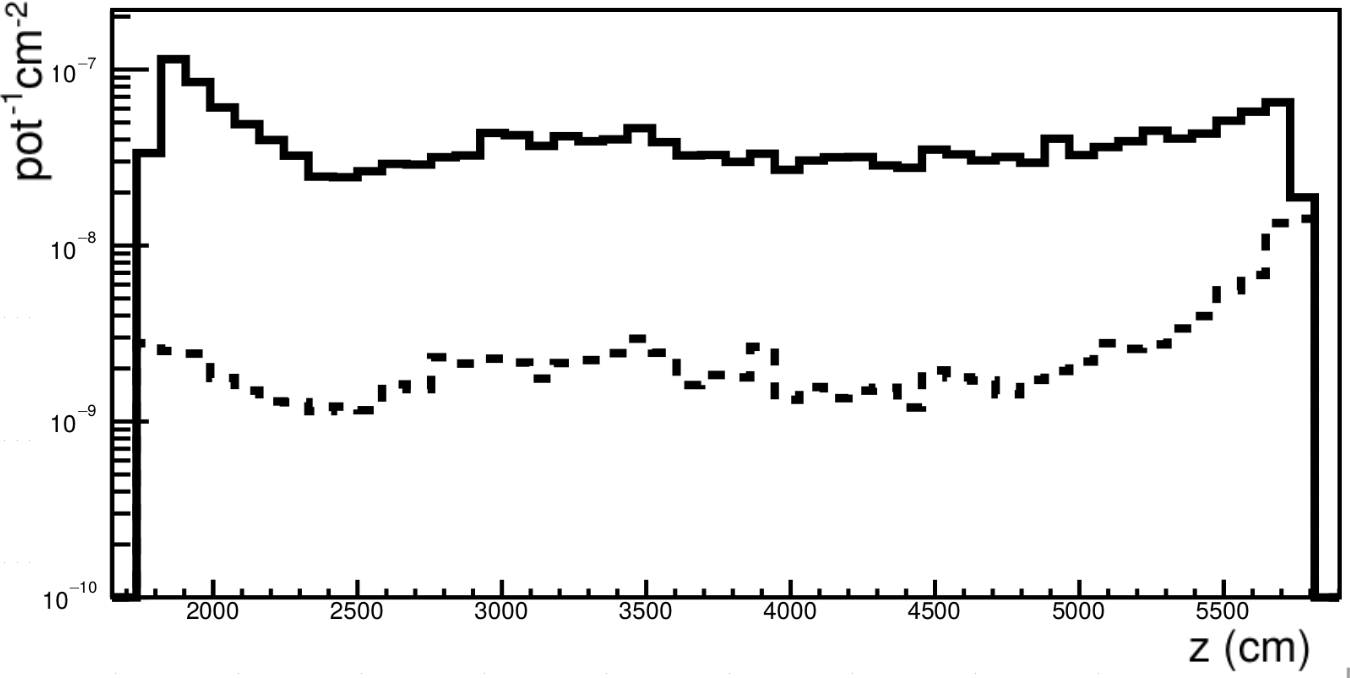}
  \caption[]{Left: Layout of the neutron shielding above the LCMs. Right: neutron reduction induced by the borated polyethylene shielding vs the longitudinal position in the tagger. The solid line represents the neutron
  flux at the inner surface of the tagger while the dashed one the flux just outside of the shielding.}
  \label{fig:neutronS}
\end{figure}

The calorimeter has been surrounded by a shielding of Borated polyehtylene
with a thickness of 30~cm as shown in Fig.~\ref{fig:neutronS}, left.
The neutron reduction induced by adding this layer of material amounts to 
a factor of $\sim$~18, averaging over the expected energy spectrum (see Fig.~\ref{fig:neutronS}, right).

\section{Test setup at the T9 beamline}
\label{sec:experiment}

\begin{figure}[!htb]
\centering
\includegraphics[width=0.9\columnwidth]{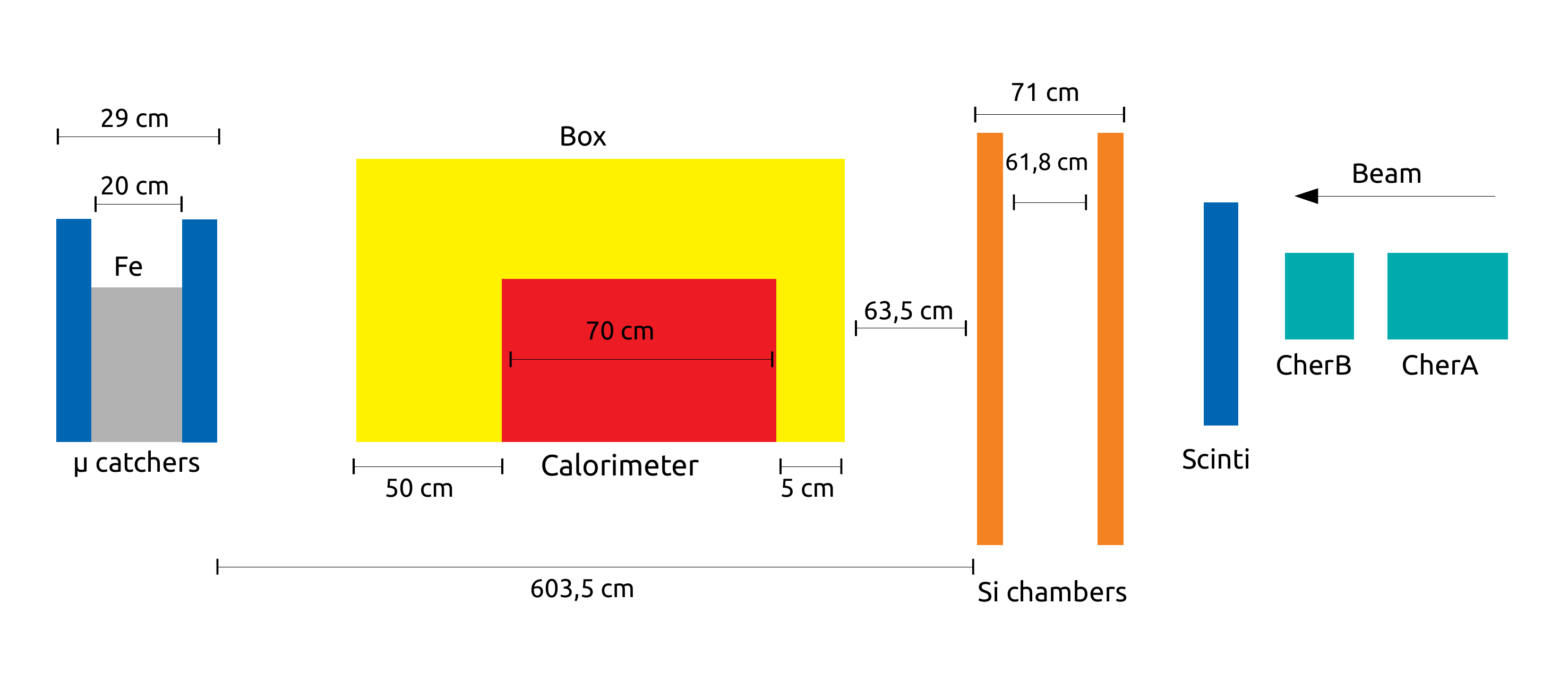}
\caption[]{\label{fig:layout_area} Layout of the instrumentation in
  the T9 experimental area. The detectors installed in T9 are the Calorimeter (Module 1,2,3), two
  Cherenkov counters (Cher A and B), two silicon chambers (Si chambers), a muon catcher ($\mu$ catchers) and the trigger scintillator plane (Scinti). Dimensions and distances are not in scale.}
\end{figure}

The calorimeter was exposed to electrons, muons and pions at the CERN PS East Area facility. We carried out a pilot run with Module 1 in May 2018 and a complete characterization of the prototype in September 2018.   The momentum of the particles was varied between 1 and 5 GeV, i.e. in the range of interest for ENUBET (1-3~GeV) and above. 
The detector was positioned inside an Aluminum box to ensure light tightness and mounted on a platform in the T9 experimental area in front of two silicon strip detectors. The layout of the instrumentation in the experimental area is shown in Fig.~\ref{fig:layout_area}.  During the data taking the calorimeter was tilted at different angles (0, 50, 100, 200~mrad) with respect to the beam direction.

The silicon detectors~\cite{Prest:2003sy,test_12module} provide track reconstruction with a spatial resolution of about 30~$\mu$m. A pair of threshold Cherenkov counters filled with CO$_2$ is located upstream of the silicon detectors. The maximum operation pressure of the counters is 2.5~bar. As a consequence, they were used to separate electrons from heavier particles ($\mu$ or $\pi$) for momenta below 3~GeV while, during runs with momenta between 3~GeV and 5~GeV, the two counters were operated at different pressures to
identify electrons, muons and pions.

The data acquisition system is triggered by a $10 \times 10$~cm$^2$ plastic scintillator located between the silicon and Cherenkov detectors. Two pads of plastic scintillator (``muon catcher'') are positioned after the calorimeter. We installed a 20~cm thick iron shield between the scintillators to select high purity samples of muons or non-interacting pions.  

Particles in the beamline are produced from the interaction of 24 GeV/c primary protons of the CERN-PS accelerator with a fixed target. As in \cite{Ballerini:2018hus}, we employed the T9 ``electron enriched'' target: an Aluminum Tungsten target (3$\times$5$\times$100~cm$^2$) followed by a Tungsten cylinder (diameter: 10~cm, length: 3~cm). The setting of the collimators was tuned to achieve a momentum bite of 1\%.
At 3~GeV the beam composition as measured by the Cherenkov counters is 12\% electrons, 14\% muons and 74\% hadrons.  We only selected negative particles in the beamline and the contamination of protons and undecayed kaons is thus negligible.

The DAQ system employed in T9 is based on a standard VME system controlled by a SBS Bit3 model 620 bridge, optically linked to a Linux PC-system. The DAQ, the digitizers, the power supply for the SiPMs and the front-end electronics for the silicon chambers are located in the proximity of the calorimeter, inside the experimental area. 
The front-end electronics also perform zero suppression in the silicon chambers~\cite{Prest:2003sy}.
The HV settings for the Cherenkov counters and the scintillators, and the configuration setting for DAQ (start-stop of the run, quality control) are performed by a PC in the Control Room connected to the main PC of the DAQ through a Gigabit Ethernet link. Users in T9 are served by a dedicated slow extraction of the protons from the PS to the target. The acquisition is hence triggered by the coincidence of the proton beam spill (duration: 400~ms) and the signal in the plastic scintillator. 
The signals from the Cherenkov counters and muon catcher are recorded for each trigger and used off-line for particle identification.   The average particle rate recorded by the DAQ during the beam-test was $\sim$500 particles per spill. This rate is limited by the throughput of the waveform digitizers in the DAQ configuration employed for the beam-test. The  maximum sustainable rate of the calorimeter and $t_0$-layer is dominated by the recovery time of the SiPMs and is $\mathcal{O}(100 \mathrm{MHz})$ per LCM.

\section{Signal equalization and response to minimum ionizing particles}
\label{sec:mip}

\begin{figure}[!htb]
\centering
\includegraphics[width=0.9\columnwidth]{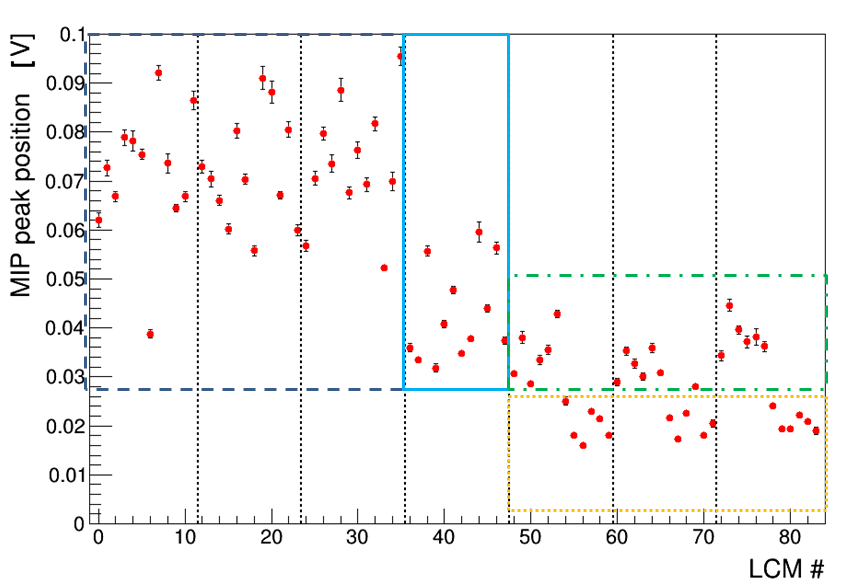}
\caption[]{\label{fig:UCM_response} Signal response to minimum ionizing particles (mip) of each LCM.
The LCMs belonging to the same Module are contained in boxes. Blue (dashed line): Module 3 with Y11 fibers. Light blue (continuous line): Module 3 with BCF92 fibers. Green (dashed line) Module 2. Orange (dotted line): Module 1.}
\end{figure}

The signal response to minimum ionizing particles (mip) of each LCM was measured using a dedicated high statistics sample of non-interacting muons and pions at 4~GeV. The bias voltage for the SiPMs was V$_{bias}$ = 32~V.
Mips are identified by Cherenkov counters located upstream of the beamline and they are selected projecting the information of the silicon chambers at the entrance of the calorimeter and checking if the particle hits a squared region of $1\times1$~cm$^2$ in the centre of each LCM.
The distribution of the signal response for each LCM is shown in Fig.~\ref{fig:UCM_response}.
The mip peak corresponds to the most probable value of the Landau fit of the deposited energy and it was 
used to equalize the relative response of the entire prototype.

\begin{figure}[!htb]
\centering
\includegraphics[width=0.6\columnwidth]{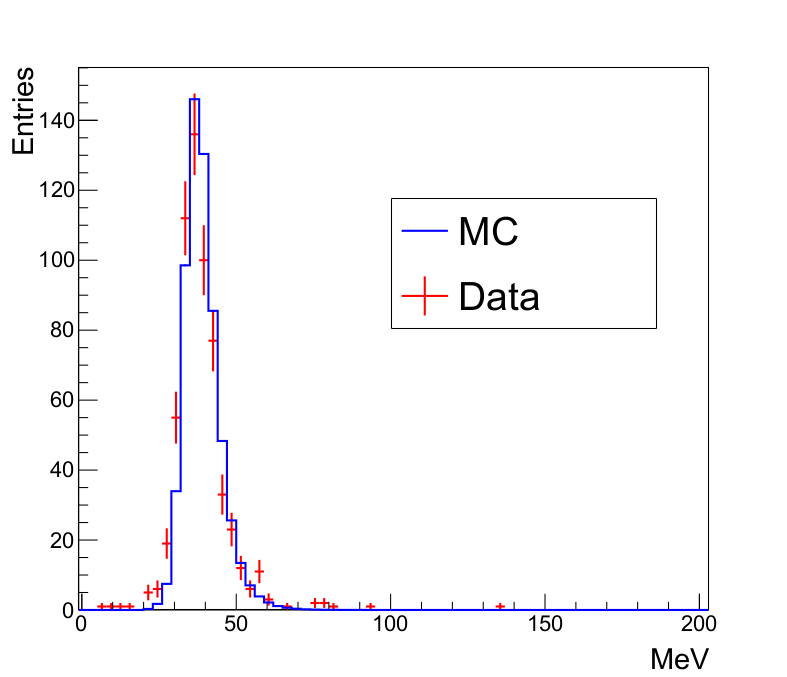}
\caption[]{\label{fig:mip}  Distribution of the energy deposited in the scintillator by 3~GeV muons impinging on the front face of the calorimeter for data (red dots) and simulation (blue line).}
\end{figure}

\begin{figure}[!htb]
\centering
\includegraphics[width=0.7\columnwidth]{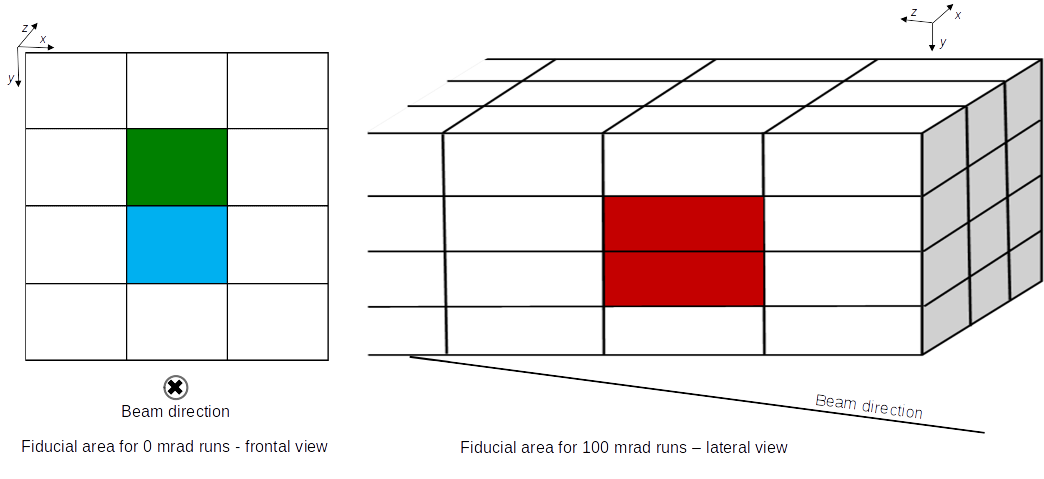}
\caption[]{\label{fig:fiducial} Fiducial areas  selected for the test beam data analysis obtained projecting the tracks reconstructed by the silicon chambers. Left: area selected for muon and pion (green square) and for electron (blue and green squares) responses on the front face of the calorimeter. Right: the red area represents the fiducial area selected for 100~mrad runs, where particles impinge from the lateral side of the inclined calorimeter. }
\end{figure}

Fig.~\ref{fig:UCM_response} shows significant variations among the LCMs.
The variation of the signal response among Modules is due to the two different optical fibers (Y11 and BCF-92) used in the assembly of the calorimeter: as expected, LCMs equipped by Y11 fibers have a higher response with respect to modules equipped with BCF-92 fibers because of the better spectral match with the plastic scintillator. In addition, the variations between Module 1 and 2 are due to the improvement of the fiber polishing procedure performed in summer 2018 (see Sec.~\ref{subsec:des_calo}). Variation of the LCMs inside the same Module were investigated by a dedicated campaign with a pulsed laser and cosmic rays. They were caused by differences in the gain of the SiPMs at 32~V. This effect was not noticed at the time of the construction  because we were not aware of the production batches of the SiPMs and the general feature of the SiPMs ($V_{bk}$ and I-V curve in reverse bias) were rather uniform. Later on, we were able to trace that the SiPMs used for the construction of the prototype were coming from different production batches but no fast equalization of the gain was possible during the beamtest  because there is not a straightforward correlation between the gain at a given voltage and the size of the reverse current. After the test, however, most of the SiPMs of Module 3 were measured one by one with a laser system and the gain difference was computed for the voltage used at CERN. Accounting for this effect, non-uniformities inside Module 3 are reduced to $\sim$10\%.    
Finally, the low signal response of 7$^{th}$ LCM of the uppermost Module (Module~3, see LCM\# 7 in Fig.~\ref{fig:UCM_response}) was traced back to an optical fiber damaged during the installation.\\
The detector response was simulated with GEANT4. 
The simulation includes the iron-scintillator tiles, the WLS fibers and a plastic box that holds all calorimeter components. 
It does not include the scintillation process and light propagation. The physics list employed is FTFP\_BERT\_HP~\cite{Allison:2016lfl,Dotti:2011jka}.
The expected signal in each LCM is thus proportional to the energy deposit in the scintillator smeared with the contribution due to photoelectron statistics (the measured value is 82~p.e./mip in a single LCM at V$_{bias}$ = 31~V).
Unlike electrons (see Sec.~\ref{sec:electrons}), saturation effects in the SiPMs are negligible. 
The mip energy deposit is in good agreement with simulations.
Fig.~\ref{fig:mip} shows the shape of the energy deposit for muons impinging on a 3 $\times$ 3~cm$^2$ on the front face of the calorimeter for data (3 GeV run, V$_{bias}$ = 31~V) and Monte Carlo simulation. This area corresponds to the green square of the Fig.~\ref{fig:fiducial}. 
Data were converted from arbitrary units (ADC counts) to MeV by equalizing the energy deposit of the electrons in the data (up to saturation) to the energy deposit of the electrons in the MC simulation.
The data response for muons is empirically reduced by 20\% to account for limitations in the MC detector description (lack of full optical simulation) and the different calorimetric response between electron and muons (e/mip ratio).

\section{Response to electrons}
\label{sec:electrons}

The calorimeter under test provides full containment of electromagnetic showers up to 5~GeV for particles impinging on the front face and from the lateral side. 
The tilted geometry reproduces its actual operating conditions in the decay tunnel, where  positrons from $K^+ \rightarrow e^+ \pi^0 \nu_e$ reach the detector with an average angle of $\sim 100$~mrad~\cite{enubet_eoi,Longhin:2014yta}.
Dedicated runs with energy (1-5~GeV) and tilt angles (0, 50, 100 and 200~mrad) relevant for neutrino physics application were carried out during the beam-test, in order to evaluate the response to electromagnetic showers.\\
Electrons were selected by the Cherenkov counters located upstream the silicon detectors. 
The bias voltage for the SiPMs was V$_{bias}$ = 31~V.
The silicon chambers are used to select single particles hitting a fiducial area with negligible lateral leakage:
for the front face run it amounts to 6$\times$3~cm$^2$ in the center of the calorimeter, while for tilted runs it corresponds to particle impinging on a projected area in the front face of the calorimeter of $0.5 \times 6$~cm$^2$, as shown in Fig.~\ref{fig:fiducial}.

\begin{figure}[htp]
\centering
\includegraphics[clip,width=0.8\columnwidth]{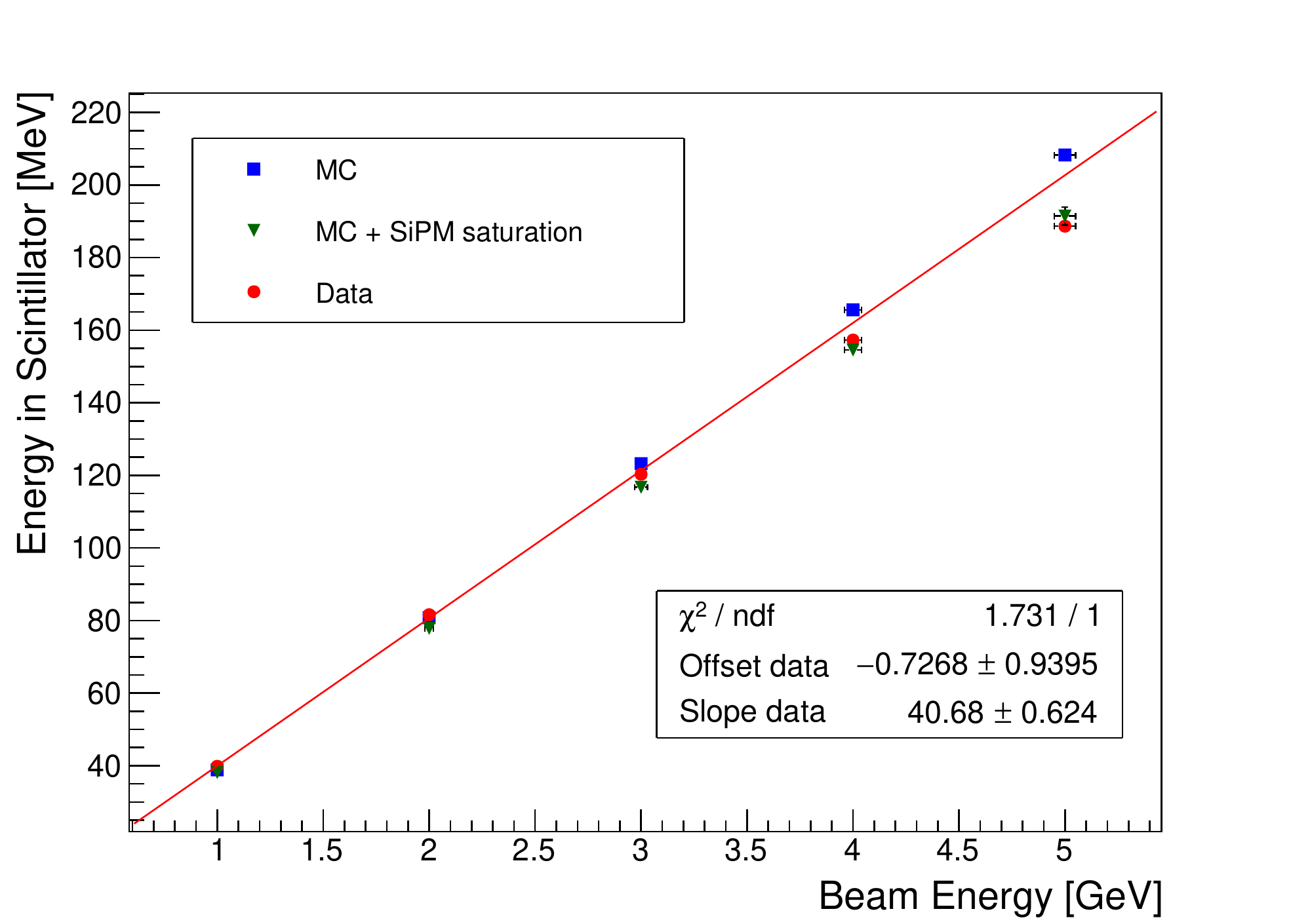}%
\caption{\label{fig:linearity} Energy reconstructed in the calorimeter versus beam energy for a 100~mrad run. Testbeam data (red dots) are compared with Monte Carlo simulation including (green triangles) and not including (blue squares) the SiPM saturation.
The horizontal errors correspond to the momentum bite of the beam. The vertical error bars (not visible in the plot, since of $\mathcal{O}$(0.1\%) and covered by the marker) in ``MC'' and ``Data'' are given by the standard error of the mean of the gaussian fit performed on the electron peaks. 
The vertical error bars in ``MC + SiPM saturation'' are given by the uncertainty on the number of pixels available to the light collection (the lowest estimate is $\sim$4580, 
while the highest estimate is $\sim$5400).}
\end{figure}

The results indicate that, as for the shashlik design~\cite{Ballerini:2018hus}, the performance of the calorimeter is the same for front and inclined runs. 
On the other hand, a clear deviation from linearity is visible above 3~GeV for all runs. Fig.~\ref{fig:linearity} shows the reconstructed energy in the scintillator for data and MC. 
The linear fit (red line) results from the data points up to 3 GeV.
At 4~(5)~GeV, the data show a deviation from linearity of $\sim3\%$ ($\sim7\%$). Most of this effect has been traced back to saturation of the SiPMs, which is enhanced by a rather large correlated noise (cross-talk) of the SiPMs at $V_{bias}=31$~V. The correlated noise was measured in a dedicated setup at the INFN Bologna labs in 2019 and turns out to be $P_{x-talk}=$ 44\% at $V_{bias}=31$~V and $P_{x-talk}\simeq$ 65\% at $V_{bias}=32$~V. To account for these effects, the average expected number of p.e. (including the single photon efficiency of the SiPM) in a LCM hit by an electron is smeared for Poisson fluctuations ($N_{pe}$) and increased by cross-talk effects: $N_{seed} \equiv (1+P_{x-talk})\cdot N_{pe}$. Note that such correction is just an approximation that neglects the correlation between saturation effects and cross talk: the latter is 
suppressed especially at high light intensities because the pixel occupancy is already nearly 100\%. For a complete treatment -- which is outside the scope of this paper -- see \cite{Gruber:2013jia,Kotera:2015rha,regazzoni2017,Acerbi:2019qgp}. In the present case, the SiPMs have 9340 cells but the fibers are put in mechanical contact with the SiPMs and illuminate a maximum number of cells $N_{max} \simeq 5000 < 9340$. We can thus approximate the number of expected fired cells in the SiPM due to saturation as
\begin{equation}
    N_{fired} \simeq N_{max} \left( 1- e^{-N_{seed}/N_{max}} \right)
    \label{eq:saturation}
\end{equation}
The uncertainty in this formula arises from the uncertainty on the actual size of the surface illuminated by the fibers ($N_{max})$ and by the above-mentioned approximations \cite{Gruber:2013jia,Kotera:2015rha}. Still, Eq.~\ref{eq:saturation} is able to account for non linearities in the detector to at least 4~GeV. Fig.~\ref{fig:linearity} shows the Monte Carlo prediction before and after the corrections for saturation effects.

\begin{figure}[!htb]
\centering
\includegraphics[width=0.9\columnwidth]{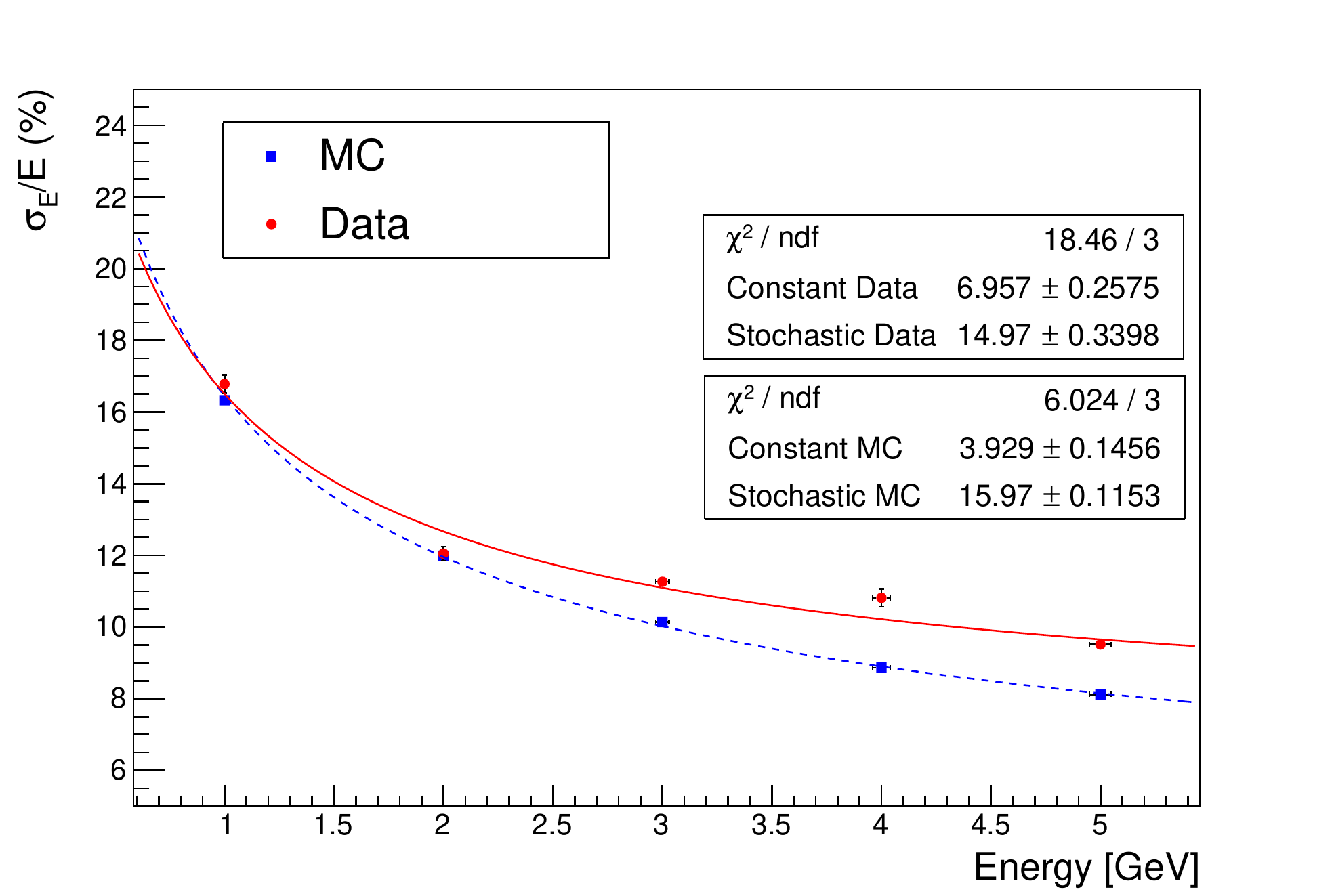}
\caption[]{\label{fig:energy_resolution}  Energy resolution versus beam energy for particles impinging on the front face (0~mrad run) for data (red dots)
and simulation (blue squares). The fit parameters for data and simulation (MC) are shown in the top and bottom insets.}
\end{figure}

The energy resolution as a function of the beam energy for particles hitting the front face of the calorimeter is shown for a 0~mrad run in Fig.~\ref{fig:energy_resolution} 
for data (red dots) and simulation (blue squares). The points are fitted to $\sigma_E/E = S/\sqrt{E
  \mathrm{(GeV)}} \oplus C $, $S$ and $C$ being the sampling
(stochastic) and constant term, respectively. 
The resolution at 1~GeV is 17\%. As expected, discrepancies with the simulation are visible in the high energy range due to the large difference in the response of the downstream modules and to SiPM saturation effects. In particular, saturation is stronger for electrons impinging in the center of the LCM where the energy is deposited mostly in a single module. Events close to the border of two adjacent modules share the deposited energy among multiple LCMs without saturating the SiPMs. This effects creates a spurious dependence on the impact point that broadens the energy distribution contributing to a worse $\sigma_E/E$
and a poorer quality of the fit. This source of non gaussianity contributes to the constant term together with standard calibration effects. In our case, calibration effects mostly results from having chosen modules and submodules (last plane of Module 3) with different components (see Tab.~\ref{tab:components}) for R\&D and procurement reasons and installing the lowest performance modules downstream the calorimeter. Hence, showers leaking after the third longitudinal plane of Module 3 are affected by lower response LCMs where, in addition, the equalization is more complex due to multiple scattering in Module 3.

\section{Response to charged pions}
\label{sec:hadrons}

Longitudinal segmented calorimeters are employed for electron/hadron separation. In ENUBET, this feature is needed to separate positrons from charged pions in the few GeV range.
The prototype under test allows for a complete longitudinal containment of pions and partial containment in the transverse direction. The response of the detector can therefore be used to validate the ENUBET simulation and the expected monitoring performance of the decay tunnel instrumentation. 

Since all variables employed by the ENUBET analysis are based on the energy deposition pattern in the LCM, this pattern was tested with a $\pi^-$ beam in the same energy range as for the electron. For this study we recorded front runs in the fiducial area depicted in Fig.~\ref{fig:layout_area} (green square).

\begin{figure}[!htb]
\centering
\includegraphics[width=\columnwidth]{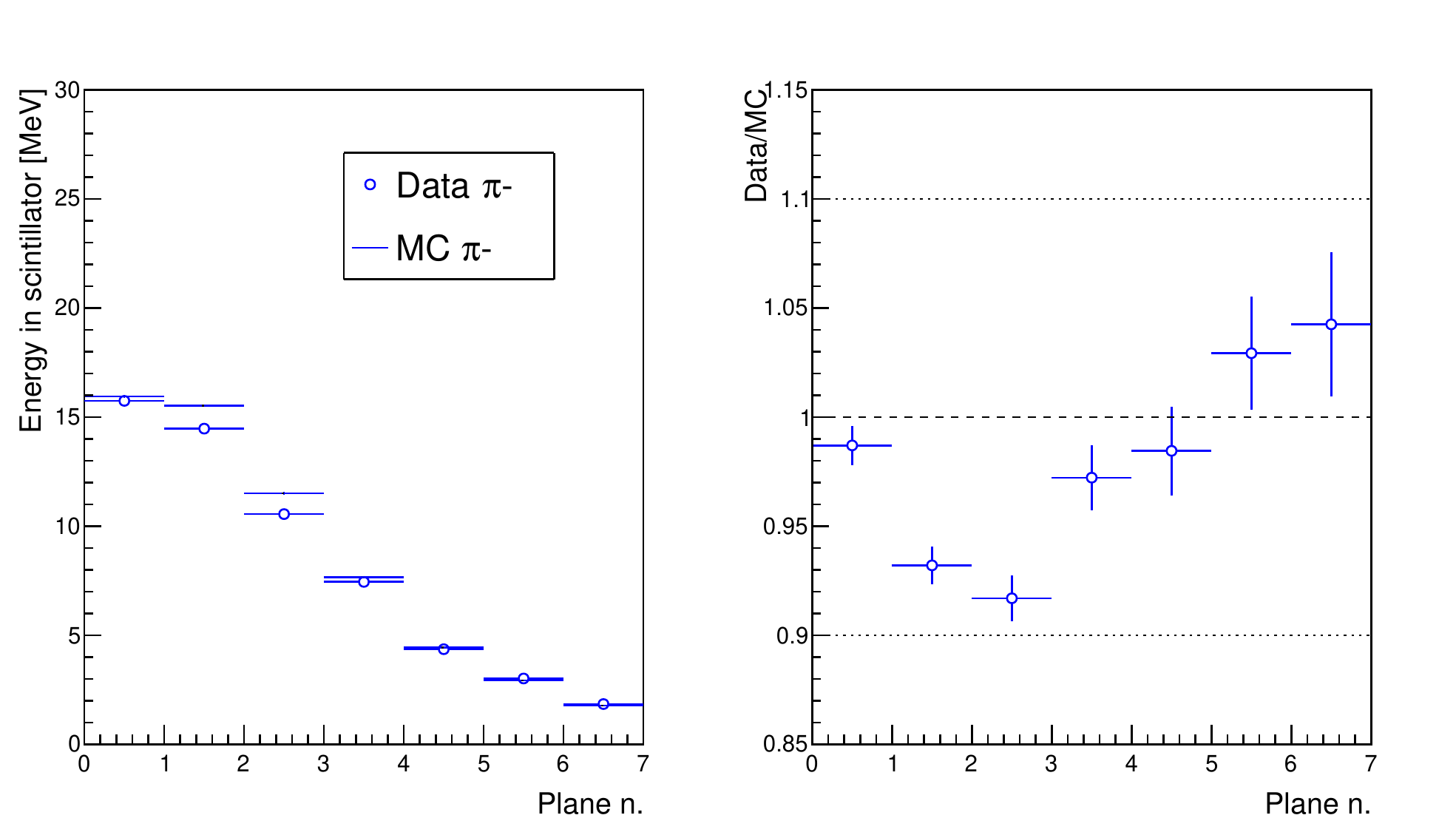}
\caption[]{\label{fig:profile_pi}
Left: average energy deposited in the scintillator as a function of calorimeter planes for 3~GeV pions. Each LCM corresponds to 0.45~$\lambda_0$. Right: energy ratio between data and MC.}
\end{figure}

Pions are selected with the Cherenkov counters (no signal in any of the counters) and traced down to the front face of the calorimeter.  
The mean $\pi^-$ energy deposited in each plane of the calorimeter is evaluated and compared with the simulation.
In this case, saturation effects are negligible and we observe no difference between the saturation-corrected Monte Carlo and the uncorrected simulation.  
Fig.~\ref{fig:profile_pi} (right) shows the average energy deposited in the scintillator (data/MC ratio) as a function of shower depth for 3~GeV pions.
The shower depth is expressed as number of calorimeter plane: plane 0 represent the front face of the calorimeter, while plane 7 is the end of the calorimeter.
The whole calorimeter
depth corresponds to $7 \times 4.3 X_0 = 30.1$~$X_0$ and
3.15~$\lambda_0$.

The data-Monte Carlo comparison is rather good:  
Fig.~\ref{fig:profile_pi} (right) shows that discrepancies do not exceed 10\% and are comparable to the uncertainty due to low-energy hadronic shower simulation~\cite{tilecal}. 

\section{Tests of the photon veto}
\label{sec:t0layer}
The $t_0$-layers were characterized in terms of single mip response, timing resolution and light collection efficiency. Moreover, to tag positrons from $K^+$ decays and reject $e^{\pm}$ pairs produced in the conversion of photons in the $t_0$-layer
the capability of the $t_0$-layer to separate one mip from two mips was investigated.

Firstly, 3 doublets were exposed to charged particles in a standalone configuration, i.e.  without the calorimeter. The trigger was given by the coincidence of two 3$\times$3 cm$^2$ scintillator pads  and a 15$\times$15 cm$^2$ pad  readout by fast Hamamatsu R9880 U-210 PMTs \cite{hama}. The SiPMs were operating with +4~V overvoltage and the signals were sampled every 0.5~ns.

\begin{figure}[t]
\centering
\subfloat[][\emph{}]
{\includegraphics[width=.48\textwidth]{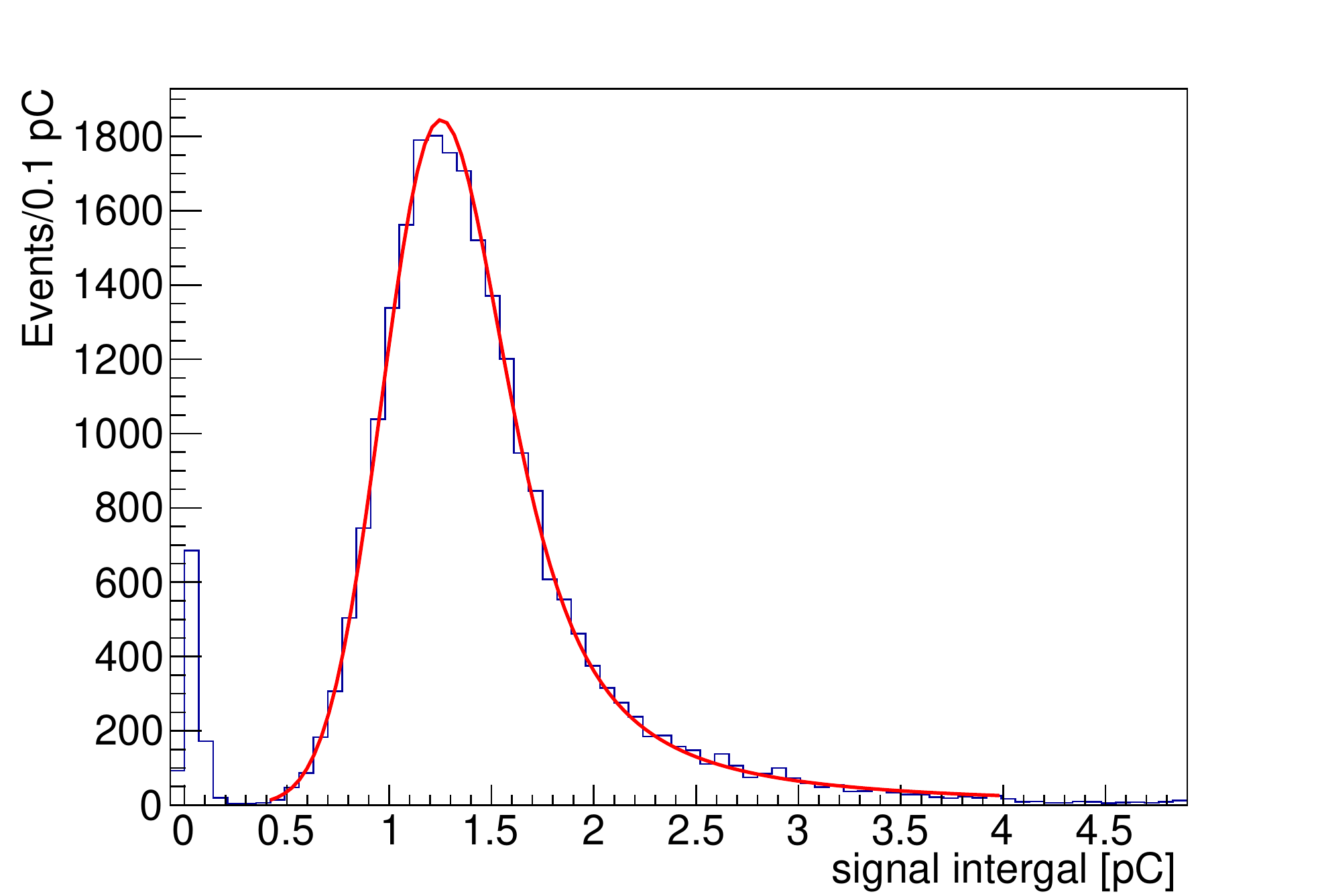}} \quad
\subfloat[][\emph{}]
{\includegraphics[width=.48\textwidth]{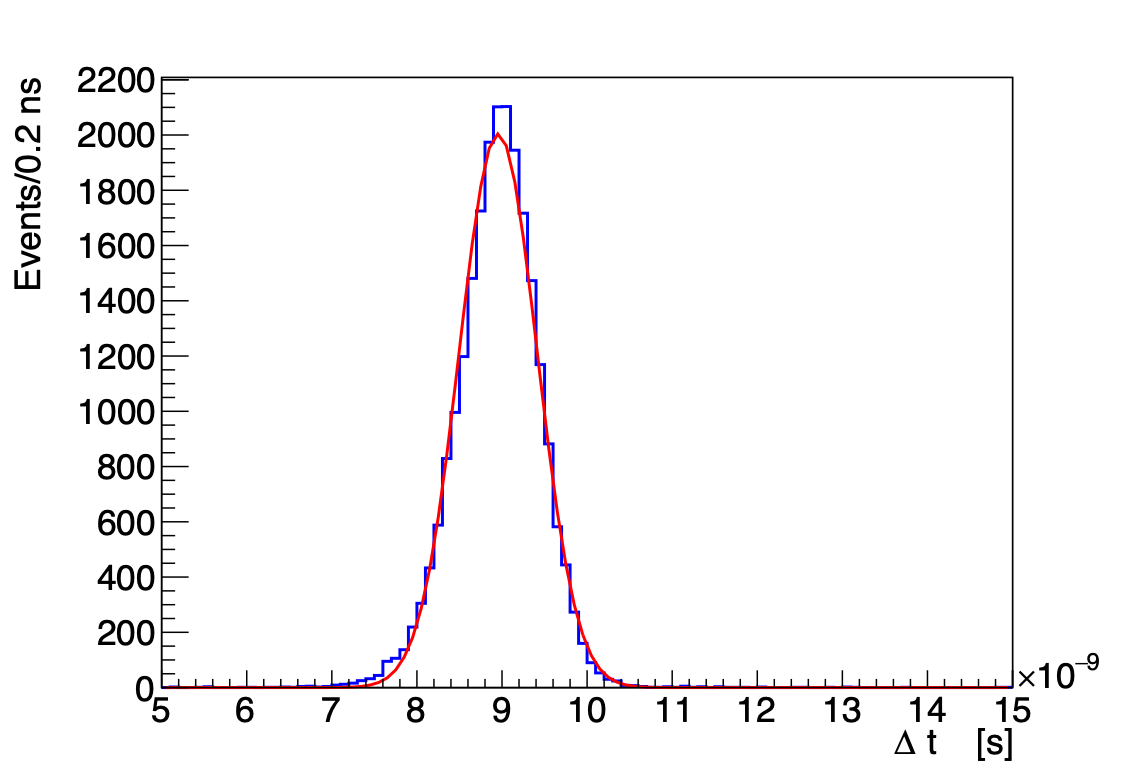}}
\caption{ (a) Signal integral distribution of the fast output in a $t_0$ tile exposed to 4~GeV pions. The distribution is fitted with a Landau convoluted with a Gaussian function. The MPV corresponds to 25~p.e. (b) Distribution of the time difference $\Delta$t at 10\% of the signal amplitude between the PMT and the SiPM signals for the fast output. The time resolution is estimated with a Gaussian fit (red line).  }
\label{fig:t0}
\end{figure}

In the standalone configuration, the $t_0$-layer tiles were exposed to 4 GeV pions. In total $\sim$~28000 1-mip events were recorded. The distribution of the integral of the fast output signals in one $t_0$-layer tile is shown in Fig.~\ref{fig:t0} (a) and is fitted with a Landau function convoluted with a Gaussian function. The MPV of the signal integral obtained from the fit corresponds to 25~p.e. for a mip crossing a single tile.

The time resolution was computed from the distribution of the time difference \it$\Delta$t \rm between the PMT and the SiPM. The threshold was set to 10\% of the signal amplitude and the time resolution is defined as the sigma of the $\Delta$t distribution. Fig.~\ref{fig:t0} (b) shows the distribution of the fast output signals from the  1 mip data. The distribution is fitted with a Gaussian function and the sigma is 460~ps. After subtracting the PMT time resolution ($\sim$ 200~ps) the measured time resolution is $\simeq$400~ps.

\begin{figure}[!htb]
\centering
\includegraphics[width=0.7\textwidth]{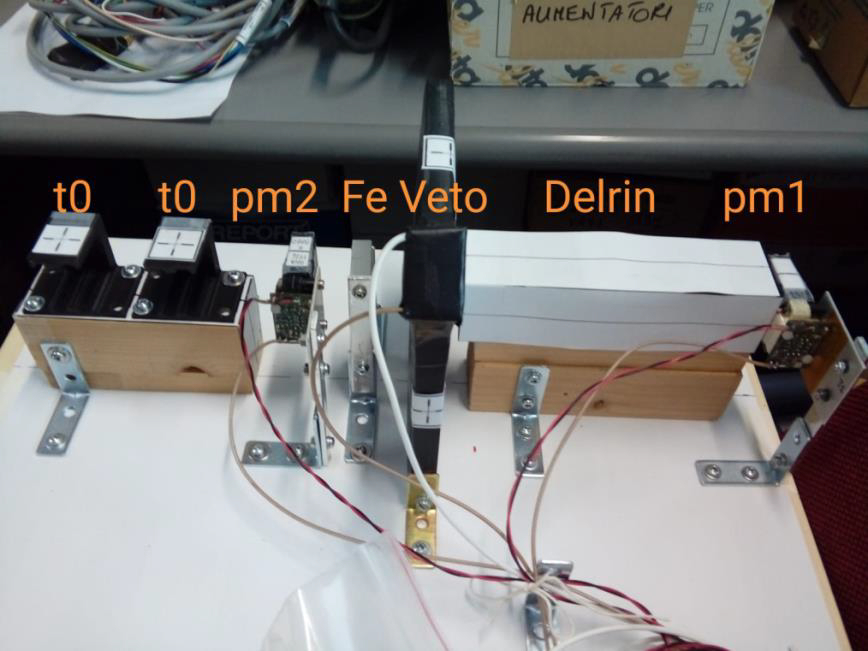}
\caption[]{\label{fig:charge_exchange} Experimental setup for the selection of charge exchange events with converted photons. ``pm1'' and ``pm2'' are the 3$\times$3 cm$^2$ scintillator pads. VETO is the 15$\times$15 cm$^2$ pad used to veto charged particles produced after the Delrin block. The trigger is given by the coincidence of pm1 and pm2 and the anti-coincidence of VETO. }
\end{figure}

\begin{figure}[t]
\centering
\subfloat[][\emph{}]
{\includegraphics[width=.41\textwidth]{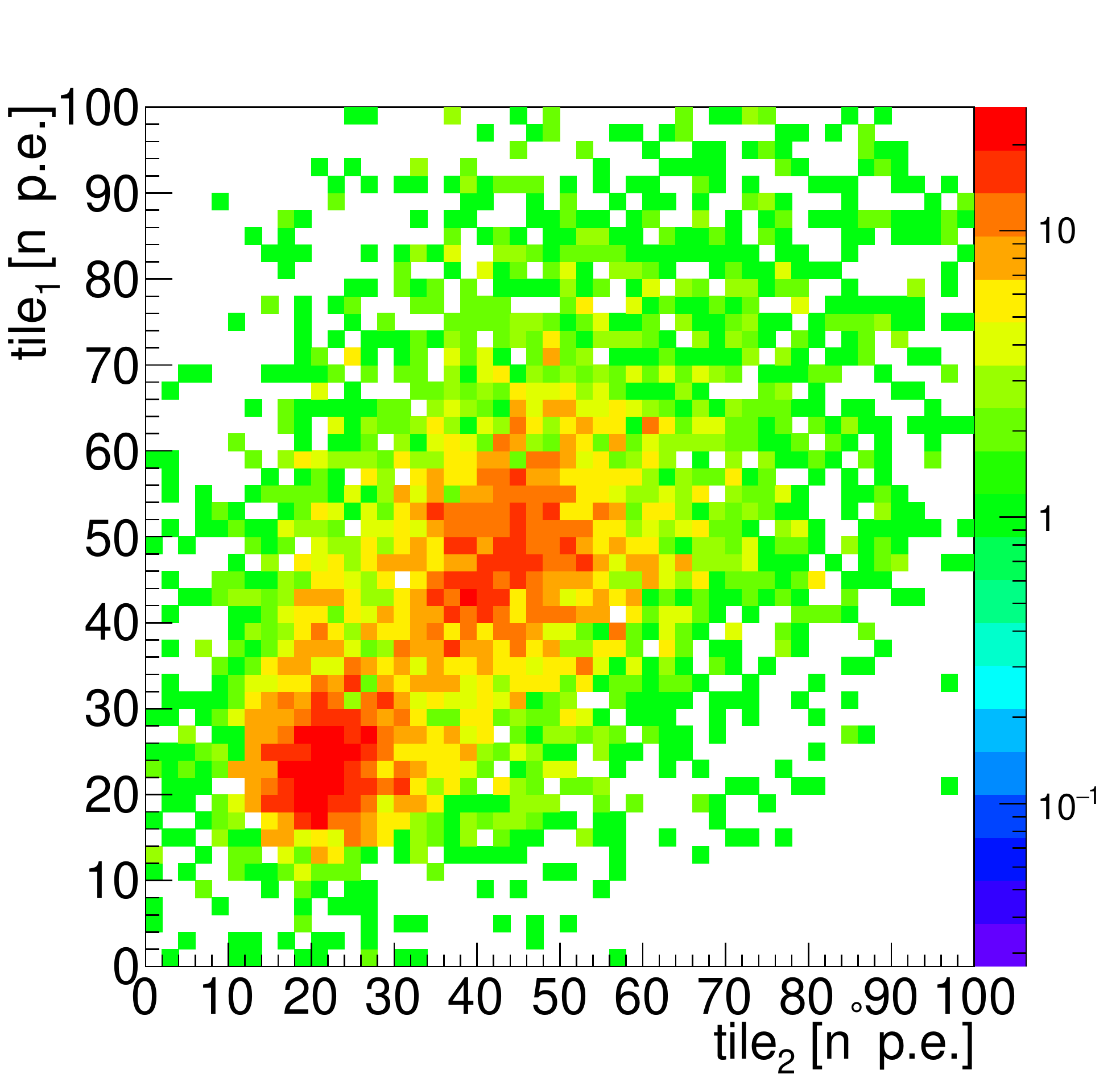}} \quad
\subfloat[][\emph{}]
{\includegraphics[width=.56\textwidth]{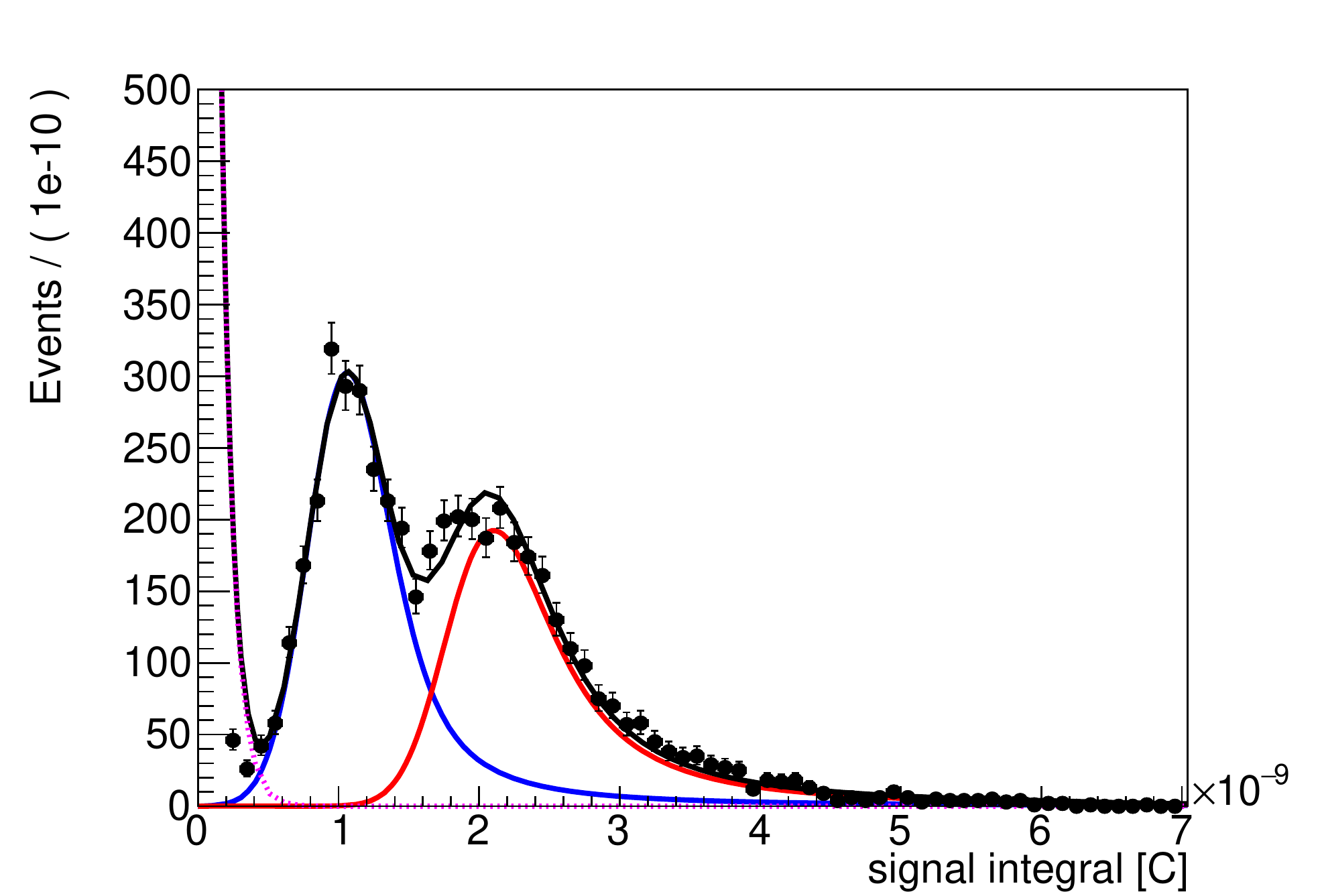}}
\caption{ (a) The number of p.e. collected in one tile versus the number of p.e. collected in another tile. (b) The fit of the data using the composite model: black dots represent the data sample, the black line is the composite model, the purple line is the dark current contribution, the blue and red lines are the signal and background pdf respectively.  }
\label{fig:12mip}
\end{figure}

The 1 versus 2 mip separation capability of the photon veto was studied exploiting charge-exchange reactions ($\pi^- + X \rightarrow \pi^0 + Y \rightarrow \gamma + \gamma + Y$) as a source of photons. We set up a new configuration (see Fig.~\ref{fig:charge_exchange}) where a block of polyoxymethylene (Derlin\textsuperscript \textregistered ) was installed along the beamline just after the 3$\times$3 cm$^2$ scintillator pad (pm1) to produce pion charge exchange.
Another 3$\times$3 cm$^2$ scintillator pad (pm2) is placed just before the $t_0$-layer. 
In addition, we installed a block of iron between the 15$\times$15~cm$^2$ scintillator pad and pm2, to convert $\gamma$ into $e^+$ $e^-$ pairs.
In this configuration the 15$\times$15~cm$^2$ pad acts as a VETO for 1 mip particles and we acquired about 29000 events. These events are a mixture of 1 mip and 2 mip events. 1 mip events are due either to the inefficiency of the VETO or to converted photons where only one charged particle reaches the $t_0$-layer. Fig.~\ref{fig:12mip} (a) shows the number of p.e. collected in one tile versus the number of p.e. collected in another tile. The two peaks corresponding to 1 and 2 mip distributions respectively are clearly visible.

In order to assess the 1 versus 2~mip separation capability the data of a single tile were used to tune a model by composing the 1 mip signal pdf (Landau convoluted with a Gaussian) with the 2 mip background pdf (Landau convoluted with another Landau convoluted with a Gaussian) together with the dark current distribution (exponential) and subtracting the noise pedestal. Fig.~\ref{fig:12mip} (b) shows the fit of the data (black dots) with the composite model (black line): the signal (blue line) and the background (red line) pdf are shown separately.

The signal and background pdf were used to generate 10$^6$ MC events of 1 mip and 2 mip crossing a tile of $t_0$-layer. A cut on the signal integral in one tile was studied assuming a signal to noise ratio N$_s$/N$_b$ of 2.3 as predicted by the MC simulation. 

The optimal cut on the signal integral that maximizes the significance is 1.8~pC for which a signal selection efficiency of 87\% and a background rejection efficiency of 89\% are obtained. The corresponding value for the purity is 95\%. 

Using the beam-test data we also studied the possibility to reduce the sampling rate down to 500 MS/s (250 MS/s). In this case, the waveforms of the fast output sampled by the digitizer every 0.5~ns were sub-sampled off-line every 2~(4)~ns. The fast output signal has a rising time of about 3.5~ns and a resolution of 500 MS/s still allows to sample the rising edge of the waveform.  
The sampling at 2~ns, hence, is suitable to measure the rising edge of the signal with a time resolution $<1$~ns, i.e. well below the requirements of ENUBET. On the other hand, sampling at 4~ns is compatible with the rising time of the signal and the waveform cannot be safely reconstructed. 
  
The light collection efficiency was measured in a configuration of 4 doublets of $t_0$-layers combined with the calorimeter (Fig.~\ref{fig:veto}). In this case, we recorded only runs with the calorimeter tilted at 100~mrad with respect to the beam axis and hence only 4 tiles were crossed by charged particles.  
Muon tracks were selected by the Cherenkov detectors and we employed the silicon chambers to identify events where the charged particle crosses the $t_0$-layer. Considering three consecutive tiles of the photon veto, we computed the efficiency of the tile located in the middle 
The efficiencies measured for the tiles under test were all above 99\%.

\section{Conclusions}
\label{sec:conclusions}

In this paper, we presented the construction and testbeam performance of a small scale prototype of the ENUBET instrumented decay tunnel. The scintillation light produced in the calorimeter and photon veto is read out by WLS fibers running along the edges of the tiles to reduce potential radiation damage of the SiPMs. The calorimeter response to mip, electrons and pions is in good agreement with expectations in the energy range of interest for ENUBET (1-3 GeV) but some discrepancies in the electron response were observed above this range (>4 GeV, see Sec.~\ref{sec:electrons}). The electromagnetic resolution is 17\% at 1 GeV and the sampling term is the dominant contribution in the 1-3 GeV range. Non-linearities in the electron response are visible at higher energies due to partial saturation of the SiPMs and non-uniformity of response of the LCMs. Both can be improved by tuning the size of the SiPMs and the equalization of the gain (see Sec.~\ref{sec:mip} and \ref{sec:electrons}). 
The photon veto was tested in standalone mode and combined with the calorimeter during the experimental campaign. The 1-mip efficiency of two doublets is $>99$\% and the 1 mip sample can be disentangled from the 2-mip component with a 95\% purity in the background conditions of ENUBET. 

In conclusion, the lateral readout calorimeter equipped with the $t_0$-layer fulfills the specifications of ENUBET and is well suited for the instrumented tunnel of monitored neutrino beams.    

\acknowledgments
This project has received funding from the European Union's Horizon 2020 Research and Innovation programme under Grant Agreement no. 681647 and the Italian Ministry for Education and Research (MIUR, bando FARE, progetto NUTECH).  The authors gratefully acknowledge CERN and the PS staff for successfully operating the East Experimental Area and for continuous supports to the users.  We thank J. Bernhard, L. Gatignon, M. Jeckel and H. Wilkens for help and suggestions during the data taking on the PS-T9 beamline.  We are grateful to the INFN workshops
of Bologna, Milano Bicocca and Padova for the construction of
detector and, in particular, to D.~Agugliaro, S.~Banfi, G.~Ceruti,  L.~Degli Esposti, M.~Furini, R.~Gaigher, L.~Garizzo, M.~Lolli,  R.~Mazza, A.~Pitacco, L.~Ramina, M.~Rampazzo, C. Valieri and F.~Zuffa.

\end{document}